\documentclass{article}

\usepackage[main, final]{neurips_2026}

\usepackage[utf8]{inputenc} 
\usepackage[T1]{fontenc}    
\usepackage{hyperref}       
\usepackage{url}            
\usepackage{booktabs}       
\usepackage{amsfonts}       
\usepackage{nicefrac}       
\usepackage{microtype}      
\usepackage{xcolor}         
\usepackage{amsmath}
\usepackage{booktabs}   
\usepackage{multirow}   
\usepackage{graphicx}   
\usepackage{xcolor}
\usepackage{mathtools}  
\usepackage{amsthm}
\usepackage{wrapfig}
\usepackage{enumitem}

\definecolor{gaingreen}{RGB}{0,97,0}
\definecolor{lossred}{RGB}{156,0,6}
\title{Bridging the Modality Bottleneck in Pathology MIL through Molecularly Guided Feature Adaptation}

\author{%
  Yucheng Xing$^{1,2}$\thanks{Equal contribution.} \And
  Pei Liu$^{3}$\footnotemark[1] \And
  Jingying Ma$^{1}$ \And
  Ruping Hong$^{4}$ \And
  Jiangdong Qiu$^{4}$ \And
  Tianyu Liu$^{1}$ \And
  Kai He$^{1}$ \And
  Ling Huang$^{1,5}$\thanks{Corresponding author: Ling Huang (\texttt{iweisskohl@gmail.com}).} \And
  Mengling Feng$^{1}$
  \AND
  \parbox[t]{0.9\textwidth}{\centering\normalfont\small
    $^{1}$National University of Singapore \\
    $^{2}$National University of Singapore Guangzhou Research Translation and Innovation Institute \\
    $^{3}$Hunan University \\
    $^{4}$Peking Union Medical College Hospital, Chinese Academy of Medical Sciences
          \& Peking Union Medical College \\
    $^{5}$Imperial College London
  }
}

\begin{document}

\maketitle

\begin{abstract}
Multiple instance learning (MIL) is the dominant framework for whole-slide image analysis in computational pathology, typically combining a frozen patch encoder, a projection layer, and a slide-level aggregator. While encoders and aggregators have been extensively studied, the projection layer remains a largely morphology-only bottleneck. This limits endpoints such as biomarker status and survival, which are governed by a molecular state that is not fully captured by H\&E morphology. We introduce Molecularly Informed Spatial-Transcriptomic Transform (MIST), a plug-in replacement for the MIL projection layer that uses paired spatial transcriptomics only during training to construct molecular views. MIST clusters gene expression profiles into cross-modal prototypes, anchors them in the frozen foundation model feature space, and uses them to reorganize H\&E patch features according to molecular programs. It requires no transcriptomics at inference and can be inserted before standard MIL aggregators. We evaluate MIST across 23 downstream tasks and 8 MIL aggregators. MIST improves 240 of 256 configurations over the standard projection layer, with an average gain of +3.5 points, observed consistently across endpoint types: +5.2 points on survival prediction, +3.3 points on tissue subtyping, and +2.1 points on biomarker prediction. Ablations show that gene-derived prototypes contribute beyond morphology-only alternatives, while spatial, biological, and pathological analyses show that cross-modal prototype affinities capture spatially coherent molecular programs from H\&E alone. Code is available at \url{https://github.com/YuchengXing99/MIST}.
\end{abstract}

\section{Introduction}
\begin{wrapfigure}{r}{0.5\linewidth}
    \centering
    \includegraphics[width=\linewidth]{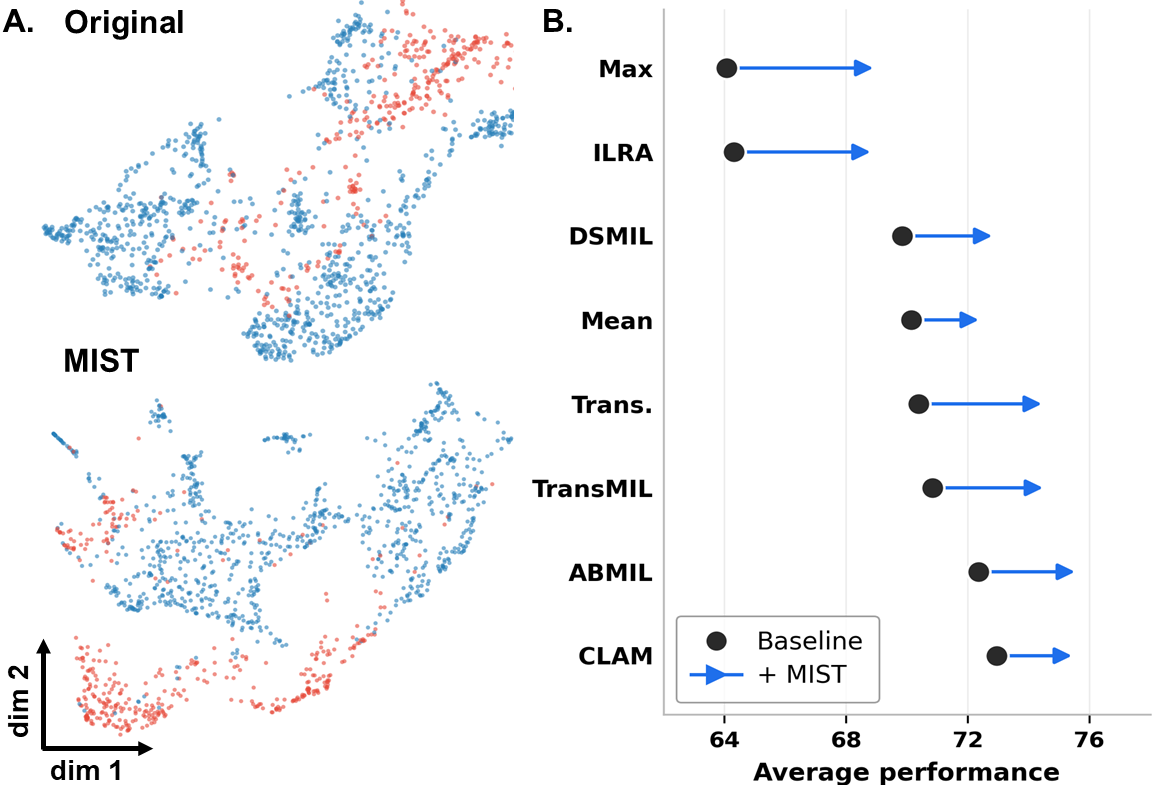}
    \caption{\textbf{(A)} MIST separates patches with similar morphology but distinct molecular profiles. Each point is an H\&E patch from HEST-1k, shown before and after MIST and colored by its matched molecular cluster.
    \textbf{(B)} Mean performance gains across downstream tasks for eight MIL aggregators.}
    \label{fig_teaser}
\end{wrapfigure}

Computational pathology (CPath) increasingly relies on whole-slide image (WSI) analysis for tissue subtyping, biomarker prediction, and prognosis~\cite{coudray2018classification,kather2019deep,chen2022pan}. Because WSIs are gigapixel-scale and heterogeneous, multiple instance learning (MIL) has become the dominant paradigm~\cite{campanella2019clinical}. A slide is split into patches, each embedded by a frozen pathology foundation model (FM), and the resulting patch set is aggregated into a slide-level prediction. This pipeline has been effective, but it exposes a fundamental mismatch. Many clinical endpoints, including biomarker status, treatment response, and survival, are governed by molecular state, whereas frozen pathology FMs are pretrained mainly from H\&E morphology and pathology reports~\cite{hoadley2018cell}. They organize tissue along morphological axes. Histologically similar regions can correspond to distinct molecular programs (Fig.~\ref{fig_teaser}A), suggesting that morphology-only MIL faces not only a capacity bottleneck~\cite{shao2026mammoth}, but also a \emph{modality bottleneck}: the projection layer lacks molecular supervision to organize H\&E features~\cite{hemker2026seal,lee2026mint,redekop2025spade}.

We address this bottleneck at the projection layer. A standard MIL pipeline contains a frozen patch encoder, a per-patch projection layer, and a slide-level aggregator~\cite{shao2026mammoth}. Prior work has extensively improved encoders through self-supervised and vision-language pretraining~\cite{ chen2024uni, lu2024visual}, and aggregators through attention, clustering, and low-rank structure~\cite{shao2021transmil,lu2021data,xiang2023exploring}. The projection layer is a strategically important intervention point because it can reshape patch features while preserving both the frozen encoder and the downstream aggregator. Existing projection-layer redesigns still operate inside the FM's morphological feature space~\cite{tang2024feature}. They can better use morphology, but receive no molecular supervision on how to organize it.

Pathologists routinely overcome this limitation by adding molecularly targeted stains. When H\&E morphology alone cannot resolve a diagnostic or prognostic question, immunohistochemistry (IHC) allows the same tissue to be interpreted under an additional molecular label space~\cite{magaki2018introduction}. Inspired by this role of staining, we introduce Molecularly Informed Spatial-Transcriptomic Transform (MIST) for MIL. MIST converts paired spatial transcriptomics (ST), available only during training, into molecular views that reorganize H\&E patch features along gene-defined axes. It leaves the pathology FM frozen, requires no transcriptomics at inference, and can be inserted before standard MIL aggregators.

Paired ST is a natural source of molecular supervision because it co-registers H\&E morphology with spot-level gene expression~\cite{jaume2024hest}. Existing training-only ST approaches usually inject this signal at the encoder stage by fine-tuning or re-pretraining a pathology FM with vision-omics objectives~\cite{hemker2026seal,xiang2026multimodal}. This validates the value of molecular supervision, but ties the resulting representation to a specific FM checkpoint and requires costly retraining when new pathology FMs are adopted. MIST instead moves molecular supervision one stage later. By operating at the projection layer, it provides an encoder-agnostic, molecularly informed adaptation module without modifying the frozen encoder. MIST is a form of cross-modal adaptation for frozen foundation-model pipelines. Instead of requiring the auxiliary modality at inference, MIST uses it during training to define a molecularly grounded coordinate system. The downstream MIL objective then learns how to use this coordinate system for each endpoint.

MIST realizes feature-level molecular views in three steps. First, it derives molecular prototypes by clustering gene-expression profiles and anchoring the resulting prototypes in the pathology FM feature space, allowing each prototype to encode the morphological signature of a distinct molecular concept. Second, each patch is softly associated with multiple prototypes through affinities, reflecting the fact that local tissue regions often express overlapping molecular programs. Finally, prototype-conditioned low-rank residual transforms generate one molecular view per molecular concept, which are combined according to the affinity weights and added back to the original patch embedding, preserving the underlying H\&E morphology as a stable reference for downstream MIL.

We evaluate MIST across 23 downstream tasks and 8 MIL aggregators using a frozen encoder. MIST improves 240 of 256 configurations over the standard projection layer, with consistent gains across endpoint types, including survival prediction (+5.2 points), tissue subtyping (+3.3 points), and biomarker prediction (+2.1 points), yielding an overall average improvement of +3.5 points. Ablation studies further show that replacing gene-derived prototypes with random or morphology-derived alternatives consistently reduces performance, suggesting that cross-modal molecular priors contribute beyond additional morphology-only modeling capacity. Our contributions are as follows:

\begin{itemize}[leftmargin=*]
    \item \textbf{A modality-bottleneck view of MIL projection layers.} We identify the projection layer as a modular but modality-limited stage in frozen-FM pathology MIL, and show that molecular priors provide an effective intervention complementary to morphology-only projection redesigns.
    
    \item \textbf{Molecular view through cross-modal prototypes.} We introduce a projection-layer mechanism that injects spatial transcriptomic priors during training while requiring only H\&E at inference, enabling molecularly informed adaptation without retraining the pathology FM.
    
    \item \textbf{Broad empirical evidence and molecular interpretability.} Across 23 tasks and 8 MIL aggregators, MIST improves most evaluated configurations (Fig.~\ref{fig_teaser}B), while spatial, biological, and pathological analyses indicate that cross-modal prototype affinities capture spatially coherent and pathologically meaningful molecular programs.
\end{itemize}

\label{intro}

\section{Related Work}
\textbf{MIL for Computational Pathology.}
The standard MIL pipeline for gigapixel WSI analysis decomposes into three stages: patch feature extraction, per-patch linear projection, and aggregation into a slide representation. Existing work focuses on the first and third stages. At the encoder level, pathology foundation models (FMs) have evolved along two directions distinguished by pretraining supervision: vision-only FMs learn patch representations through self-supervision on unlabeled H\&E~\cite{chen2024uni, vorontsov2024foundation, xu2024whole, chen2022scaling}, while vision-language FMs additionally align H\&E with pathology reports through contrastive objectives~\cite{lu2024visual, xiang2025vision, ding2025multimodal}. At the aggregation level, MIL methods pool patch sets through inductive biases such as attention~\cite{ilse2018attention}, clustering-constrained attention with instance supervision~\cite{lu2021data}, dual-stream max-pooling~\cite{li2021dual}, correlated self-attention~\cite{shao2021transmil}, and low-rank structure~\cite{xiang2023exploring}.

The intermediate projection layer has only recently received dedicated attention. Regional Transformers re-embed patches based on spatial proximity~\cite{tang2024feature}, query-aware sparse attention extends this to longer-range spatial dependencies~\cite{guo2025context}, and low-rank mixtures of experts adapt the per-patch transformation to each patch's morphological phenotype~\cite{shao2026mammoth}. These methods reorganize patches along progressively finer dimensions, moving from spatial to morphological similarity, yet all draw their inductive bias from the FM's feature space alone. MIST instead injects transcriptional prior from paired ST, introducing supervision from a modality the FM never saw, with no ST required at inference.

\textbf{Spatial Transcriptomics for Downstream Pathology Tasks.}
Recent studies have explored paired molecular and H\&E data for downstream MIL tasks such as tissue subtyping, biomarker prediction, and survival analysis. Multimodal fusion approaches combine H\&E features with bulk molecular profiles~\cite{chen2022pan,xingemms}, but require gene expression at inference, which is rarely available in clinical practice. ST offers a finer-grained alternative, providing molecular signal co-localized with morphology at spot resolution. Training-only ST approaches incorporate this signal at the encoder stage through paired-ST objectives such as vision-omics contrastive fine-tuning~\cite{hemker2026seal}, ST-token distillation under DINO self-supervision~\cite{lee2026mint}, and vision-ST mixture-of-experts pretraining on HEST-1k~\cite{redekop2025spade}. These studies demonstrate the value of training-time molecular supervision, but tightly couple the learned representations to a specific FM checkpoint. Given the rapid evolution of pathology FMs~\cite{chen2024uni, lu2024visual}, the costly vision-omics training must be repeated whenever a new FM is adopted. In contrast, MIST operates at the projection layer above any frozen FM, so adopting a new FM only requires rebuilding the prototype bank.

\textbf{Prototype-Based Methods in Computational Pathology.}
Prototype-based methods learn a compact set of representative concepts to summarize complex inputs ~\cite{huang2024evidential, huang2025evidential, huang2025esurvfusion,xingsaefs,masscope}. In CPath, prior work models patches as Gaussian mixtures and aggregates through the mixture parameters~\citep{song2024morphological}, clusters patches into histological types and represents slides as type-frequency distributions~\cite{vu2023handcrafted}, extends prototype-based representations to multimodal survival prediction with bulk transcriptomics at both training and inference~\cite{song2024multimodal}, and couples morphological prototypes with class anchors under an evidential fusion framework for uncertainty-aware survival~\citep{xing2025dpsurv}. In all of these methods, prototypes live in the FM's morphology feature space and act at the aggregation stage. MIST departs in two coupled ways: prototypes are discovered in
gene-expression space but anchored as centroids in morphology space, encoding the morphological signatures of molecular concepts; and they condition the per-patch projection rather than summarize the slide, leaving the
aggregator unconstrained.

\label{related work}

\section{Method}
\begin{figure}[t]
    \centering
    \includegraphics[width=0.95\linewidth]{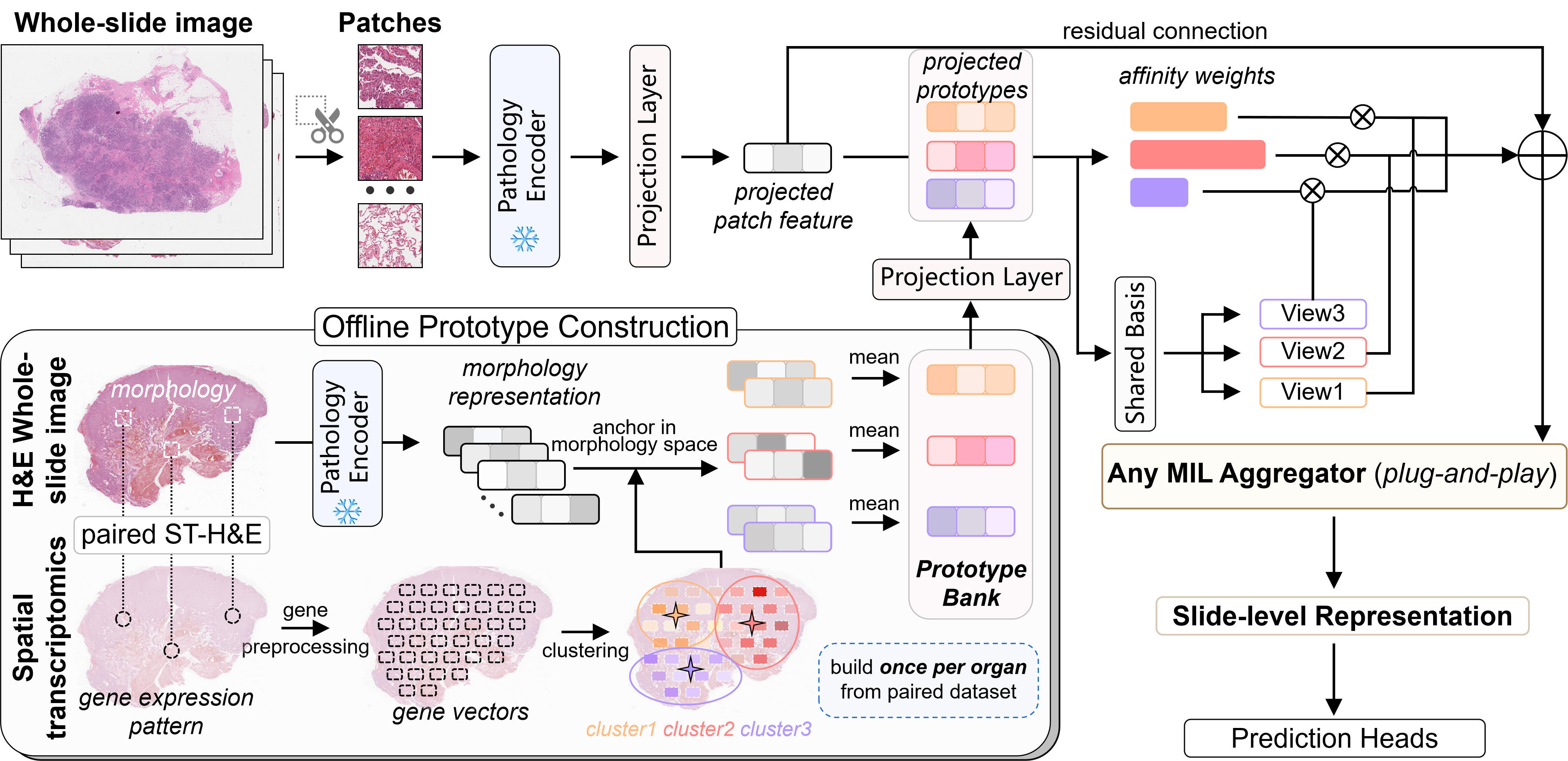}
    \caption{\textbf{Overview of MIST.}
      Each patch is encoded by a frozen foundation model and matched to organ-specific molecular prototypes through soft affinities. Prototype-conditioned low-rank residual transforms generate molecular views over a shared basis, which are combined to produce a patch representation consumable by any MIL aggregator. The prototype bank is constructed once per organ from paired ST and H\&E data, with no transcriptomics required at inference. Each prototype $\boldsymbol{\mu}_k \in \mathbb{R}^{1536}$ (bank $K{\times}1536$) is projected to $\mathbf{p}_k = \phi(\boldsymbol{\mu}_k)$.}
    \label{fig:mist_architecture}
\end{figure}

To bridge the transcriptional modality gap without requiring ST at inference, we propose \textbf{MIST} (\textbf{M}olecularly \textbf{I}nformed \textbf{S}patial-Transcriptomic \textbf{T}ransform), a plug-and-play replacement for the standard MIL projection layer that, unlike prior MIL redesigns, conditions patch transforms on gene-expression prototypes constructed offline from paired ST data (Fig.~\ref{fig:mist_architecture}).
MIST (1) builds molecular prototypes via gene-space clustering, anchored in morphological feature space for direct H\&E conditioning (§\ref{sec:prototype_construction}); (2) assigns patches via soft, non-competitive affinities to handle co-expression (§\ref{sec:affinity}); (3) applies a prototype-conditioned low-rank transform over a shared morphological subspace to produce molecular patch views (§\ref{sec:lowrank}); and (4) mixes views residually by affinity, preserving the original morphological signal as a stable anchor (§\ref{sec:mixing}).

\subsection{Offline Cross-Modal Prototype Construction from Paired ST-H\&E Data}
\label{sec:prototype_construction}

We derive the prototypes from paired ST and H\&E data in HEST-1k~\cite{jaume2024hest}, which provides gene expression and morphology co-localized at Visium-spot resolution across 1{,}276 paired slides and 26 organ types with millions of spots.

\textbf{Extracting Paired Morphology and Gene Expression at Each Spot.} HEST-1k provides pre-extracted patches at $0.5\,\mu$m/px, each centered on a Visium spot. We encode each patch with the frozen pathology foundation model UNI2-h~\cite{chen2024uni} to obtain its morphology embedding $\mathbf{x}_j \in \mathbb{R}^{D}$. For the corresponding Visium spots, we follow SEAL's processing pipeline~\cite{hemker2026seal} to obtain the gene expression vector $\mathbf{g}_j$: empty spots and genes detected in less than 10\% of spots are removed; raw counts are normalized to $10^4$ per spot and $\log(1{+}x)$-transformed; the top $2{,}000$ highly variable genes are retained via the Seurat~v3 method~\cite{stuart2019comprehensive}; and each spot's expression is averaged with the mean of its immediate neighbors on the hexagonal lattice to mitigate Visium sparsity and noise. This yields paired samples $\{(\mathbf{x}_j, \mathbf{g}_j)\}_{j=1}^{M}$. Further details are provided in Appendix~\ref{app:prototype_construction}.

\textbf{Clustering Within Each Organ to Form Prototypes.} We construct prototypes per organ since molecular concepts differ substantially across organs and downstream tasks are typically organ-specific. For each organ $o$, we cluster its gene vectors $\{\mathbf{g}_j\}$ into $K_o$ groups (selected by silhouette score), assigning each spot a cluster label $c_j$. The prototype for cluster $k$ is the centroid of the corresponding patch features:
\begin{equation}
    \boldsymbol{\mu}_k = \frac{1}{|\mathcal{C}_k|} \sum_{j \in \mathcal{C}_k} \mathbf{x}_j \in \mathbb{R}^{D}, \qquad \mathcal{C}_k = \{j : c_j = k\}.
    \label{eq:prototype}
\end{equation}
This construction is central to MIST: clustering in gene space discovers molecular concepts, while the prototypes themselves live in morphological feature space so they can condition H\&E patch embeddings directly. Each $\boldsymbol{\mu}_k$ thus encodes the morphological signature of a specific molecular concept, and defines the corresponding molecular view applied to patches downstream.

\subsection{Patch Assignment to Multiple Molecular Concepts}
\label{sec:affinity}

Given the prototypes $\{\boldsymbol{\mu}_k\}_{k=1}^{K}$, we associate each patch in a WSI with these molecular concepts. To compute the affinity of a patch to each prototype, we first project both into a shared $d$-dimensional space ($d{=}512$) via a linear layer $\phi$, yielding $\mathbf{h}_i = \phi(\mathbf{x}_i)$ and $\mathbf{p}_k = \phi(\boldsymbol{\mu}_k)$. The affinity of patch $i$ to prototype $k$ is then
\begin{equation}
    \alpha_{ik} = \frac{\sigma(\mathbf{h}_i^\top \mathbf{p}_k)}{\sum_{k'} \sigma(\mathbf{h}_i^\top \mathbf{p}_{k'})},
    \label{eq:affinity}
\end{equation}
where $\sigma$ is the sigmoid, allowing a patch to activate multiple prototypes simultaneously when it expresses several molecular concepts at once. This reflects a common situation in tissue, where a single spot can carry mixed cellular programs. The $\ell_1$ normalization preserves the relative pattern of co-activations while bounding the total affinity of each patch to a fixed budget, keeping the view intensity comparable across patches before aggregation.

\subsection{Prototype-Conditioned Low-Rank Transforms}
\label{sec:lowrank}

To represent how each patch manifests under each molecular concept, each prototype $k$ induces its own view $\mathbf{v}_{ik}$ of the patch embedding, computed as
\begin{equation}
    \mathbf{v}_{ik} = \mathbf{h}_i + \mathbf{U}_k \mathbf{V} \mathbf{h}_i,
    \label{eq:views}
\end{equation}
with $\mathbf{V} \in \mathbb{R}^{r \times d}$ shared across prototypes and $\mathbf{U}_k \in \mathbb{R}^{d \times r}$ prototype-specific ($r{=}16$). The residual form mirrors the physical nature of staining. A stained sample still contains the underlying tissue, and the stain is added on top. The factorization further restricts this stain. Morphology has only a limited number of axes along which any stain can act, so the shared $\mathbf{V}$ extracts these common axes while each $\mathbf{U}_k$ recombines them for its own concept.

\subsection{Patch Representation for Downstream MIL}
\label{sec:mixing}

Given the $K$ views $\{\mathbf{v}_{ik}\}_{k=1}^{K}$, we combine them into a single patch embedding that the downstream aggregator can consume. Each patch is transformed by the concepts it expresses, weighted by its affinity,
\begin{equation}
    \tilde{\mathbf{h}}_i = \mathbf{h}_i + \sum_{k=1}^{K} \alpha_{ik}\, \mathbf{v}_{ik}.
    \label{eq:staining}
\end{equation}
The affinity weights give each patch its own view mixture on a comparable scale across patches. Adding $\mathbf{h}_i$ once more keeps the original view of the patch available alongside the molecular views, so the original morphological signal from the frozen encoder remains accessible regardless of how the patch aligns with the prototypes.

\textbf{Theoretical grounding.}
MIST can amplify the separation between morphologically similar patches that receive different prototype affinities. For two patches with $\mathbf{h}_i - \mathbf{h}_j = \epsilon\mathbf{v}$ but different prototype affinities, the output difference decomposes as
\begin{equation}
    \tilde{\mathbf{h}}_i - \tilde{\mathbf{h}}_j
    \;=\; (2\mathbf{I} + \bar{\mathbf{W}})\,\epsilon\mathbf{v}
    \;+\; \Bigl(\sum_k \Delta\alpha_k\,\mathbf{U}_k\Bigr)
    \mathbf{V}\bar{\mathbf{h}},
    \label{eq:separation-decomp}
\end{equation}
where $\epsilon\mathbf{v}$ denotes the small morphological difference between the two patches, with $\epsilon$ controlling the perturbation magnitude and $\mathbf{v}$ the perturbation direction. The first term corresponds to the morphology-only transformation, while the second term captures molecular routing induced by differences in prototype affinity. Here, $\bar{\mathbf{W}}$ denotes the uniform-affinity baseline transform and $\Delta\alpha_k = \alpha_k(\mathbf{h}_i) - \alpha_k(\mathbf{h}_j)$ measures the affinity difference between the two patches. When the routing term aligns with the morphology term, the output distance exceeds the morphology-only baseline. A formal existence proof is provided in Appendix~\ref{app:theory}.
\label{method}

\section{Experiments}
\subsection{Datasets}
\label{sec:datasets}

\textbf{Tissue subtyping.} We evaluated MIST on six tissue subtyping classification tasks: EBRAINS fine-grained (EBRAINS-F, $C{=}30$) and coarse-grained (EBRAINS-C, $C{=}12$) brain tumor subtyping ($n{=}2{,}319$ slides)~\cite{roetzer2022digital}; Non-Small Cell Lung Carcinoma (NSCLC, $C{=}2$) subtyping with 5-fold cross-validation (CV) on TCGA ($n{=}1{,}041$), with external validation on CPTAC ($n{=}1{,}091$) and NLST ($n{=}1{,}008$)~\cite{campbell2016distinct}; ISUP grading from the PANDA prostate cancer challenge ($C{=}6$, $n{=}10{,}616$)~\cite{bulten2022artificial}; and BRACS breast carcinoma subtyping at coarse (BRACS-C, $C{=}3$) and fine (BRACS-F, $C{=}7$) granularity ($n{=}547$)~\cite{brancati2022bracs}. We report balanced accuracy for the multiclass tasks, AUROC for NSCLC, and weighted $\kappa$ for PANDA.

\textbf{Molecular biomarker prediction.} We evaluate on 13 molecular biomarker prediction tasks across four cohorts. On TCGA GBMLGG, we perform 5-fold CV for IDH1 mutation prediction (GBMLGG-C, $C{=}2$, $n{=}1{,}539$) and histomolecular subtyping (GBMLGG-F, $C{=}5$, $n{=}1{,}516$), with external evaluation on the EBRAINS glioma cohort ($n{=}894$)~\cite{roetzer2022digital}. 5-fold CV on TCGA Lung (LUAD + LUSC) for TP53, KRAS, STK11, and EGFR mutation status ($C{=}2$, $n{=}1{,}024$)~\cite{weinstein2013cancer}; 5-fold CV on TCGA BRCA for receptor status (ER, PR, HER2; $C{=}2$) and PIK3CA mutation ($C{=}2$), with $n$ ranging from $741$ (HER2) to $1{,}061$ (PIK3CA); and 10-fold CV on BCNB ($n{=}1{,}058$)~\cite{xu2021predicting} for ER, PR, and HER2 status. We report AUROC for binary tasks, balanced accuracy for GBMLGG-F, and macro-averaged performance across cohorts where external evaluation is available.

\textbf{Survival prediction.} We predict disease-specific survival (DSS) with 5-fold CV on four TCGA cohorts~\cite{weinstein2013cancer}: breast carcinoma (BRCA, $n{=}1{,}041$), colon adenocarcinoma (COAD, $n{=}417$), lung adenocarcinoma (LUAD, $n{=}456$), and lung squamous cell carcinoma (LUSC, $n{=}471$). LUAD, LUSC, and COAD
include external validation on CPTAC; LUAD and LUSC additionally include NLST. Performance is measured by the concordance index (C-index).

\textbf{Prototype source data.} Prototypes are built from HEST-1k~\cite{jaume2024hest}, which shares no patients or slides with any downstream cohort. One HEST slide per organ is held out for interpretability (Section~\ref{sec:intrep}).

\subsection{Baselines}
\label{sec:baselines}
We compare MIST against the standard linear layer across eight MIL aggregators: ABMIL~\cite{ilse2018attention}, CLAM~\cite{lu2021data}, DSMIL~\cite{li2021dual}, TransMIL~\cite{shao2021transmil}, ILRA~\cite{xiang2023exploring}, Transformer~\cite{wagner2023transformer,vaswani2017attention}, MeanMIL, and MaxMIL. Each aggregator retains its published hyperparameters. We further combine MIST with MAMMOTH~\cite{shao2026mammoth}, with an H\&E-only prototype control, to examine whether modality bottlenecks and morphology-focused capacity bottlenecks are complementary. Further details are provided in Appendix~\ref{app:baselines}.

\subsection{Implementation Details}
\label{sec:implementation}
All slides are processed at $20\times$ magnification and tiled into non-overlapping $256{\times}256$ patches. Patch features are extracted using the frozen UNI2-h pathology foundation model~\citep{chen2024uni}. Molecular prototypes are constructed once per organ from paired ST and H\&E data following Section~\ref{sec:prototype_construction}. Unless otherwise specified, all models are trained using AdamW with cosine learning rate decay and evaluated under the same optimization setting across all MIL aggregators. Survival, biomarker, and NSCLC tasks use matched cross-validation splits for Base and MIST; PANDA, BRACS, and EBRAINS use predefined splits with 1{,}000 bootstrap resamples of the test set. Additional implementation details and hyperparameters are provided in the appendix~\ref{app:implementation}.
\label{experiments}

\section{Results}
\subsection{Downstream Task Performance}
\label{sec:performance}

Across all 14 summary results in Tables~\ref{tab:survival_summary}--\ref{tab:classification_summary} (4 survival cohorts, 4 biomarker datasets, and 6 subtyping tasks), \textsc{Mist} improves the aggregator-averaged performance in all cases (mean gain $+3.5$ points, 95\% CI $[2.6, 4.4]$; one-sided Wilcoxon signed-rank test, $p < 10^{-4}$).

\textbf{Survival prediction.} Table~\ref{tab:survival_summary} shows that \textsc{Mist} improves mean concordance across all four datasets, yielding an average gain of $+5.2$ points. Table~\ref{tab:survival_detailed} further shows performance improvements in 71 out of 72 experimental configurations. The largest gains are observed for weaker aggregators that rely on coarse pooling operations. For example, MaxMIL achieves a $+7.2$-point improvement, approximately twice the typical gain obtained by stronger backbones such as ABMIL or CLAM. This trend is consistent with the hypothesis that survival outcomes are primarily driven by molecular state and microenvironmental composition, which H\&E morphology captures only indirectly~\cite{chen2022pan}. Consequently, weaker aggregators have limited capacity to fully exploit morphological information alone, causing them to benefit more substantially from molecularly-informed priors.

\begin{table}[t!]
\centering
\caption{\textbf{Survival prediction.} Concordance index across four datasets using eight MIL aggregators. LUAD and LUSC are averaged across TCGA, CPTAC, and NLST, while COAD is averaged across TCGA and CPTAC. Results use matched 5-fold CV splits for Base and MIST; standard errors across folds are in parentheses. Trans.,Transformer.}
\label{tab:survival_summary}
\resizebox{\textwidth}{!}{%
\begin{tabular}{llccccccccc}
\toprule
Cohort & Status & ABMIL & CLAM & TransMIL & Trans. & ILRA & Mean & Max & DSMIL & Avg. \\
\midrule
\multirow{3}{*}{BRCA}
 & Base  & 60.58$_{(6.44)}$ & 60.79$_{(6.33)}$ & 62.66$_{(9.88)}$ & 63.12$_{(8.92)}$ & 57.47$_{(6.20)}$ & 65.77$_{(7.26)}$ & 54.95$_{(3.82)}$ & 60.99$_{(4.92)}$ & 60.79$_{(3.14)}$ \\
 & +Ours & 69.34$_{(3.87)}$ & 69.70$_{(3.65)}$ & 67.56$_{(7.98)}$ & 70.28$_{(8.02)}$ & 65.31$_{(6.37)}$ & 70.96$_{(1.54)}$ & 58.77$_{(9.11)}$ & 68.12$_{(2.51)}$ & 67.50$_{(3.70)}$ \\
 & $\Delta$ & \textcolor{gaingreen}{+8.76} & \textcolor{gaingreen}{+8.91} & \textcolor{gaingreen}{+4.90} & \textcolor{gaingreen}{+7.16} & \textcolor{gaingreen}{+7.84} & \textcolor{gaingreen}{+5.19} & \textcolor{gaingreen}{+3.81} & \textcolor{gaingreen}{+7.13} & \textcolor{gaingreen}{+6.71} \\
\midrule
\multirow{3}{*}{LUAD}
 & Base  & 66.07$_{(6.56)}$ & 65.75$_{(6.72)}$ & 60.90$_{(5.23)}$ & 56.15$_{(4.79)}$ & 51.03$_{(3.37)}$ & 66.62$_{(6.40)}$ & 49.40$_{(4.84)}$ & 64.81$_{(5.63)}$ & 60.09$_{(6.57)}$ \\
 & +Ours & 67.49$_{(6.54)}$ & 67.33$_{(6.85)}$ & 65.64$_{(9.18)}$ & 62.32$_{(5.90)}$ & 55.94$_{(3.62)}$ & 67.64$_{(6.89)}$ & 58.74$_{(4.41)}$ & 66.37$_{(6.32)}$ & 63.93$_{(4.19)}$ \\
 & $\Delta$ & \textcolor{gaingreen}{+1.42} & \textcolor{gaingreen}{+1.59} & \textcolor{gaingreen}{+4.74} & \textcolor{gaingreen}{+6.17} & \textcolor{gaingreen}{+4.91} & \textcolor{gaingreen}{+1.02} & \textcolor{gaingreen}{+9.35} & \textcolor{gaingreen}{+1.55} & \textcolor{gaingreen}{+3.84} \\
\midrule
\multirow{3}{*}{LUSC}
 & Base  & 61.99$_{(2.75)}$ & 61.27$_{(1.81)}$ & 53.79$_{(4.86)}$ & 54.71$_{(3.01)}$ & 49.29$_{(6.68)}$ & 61.79$_{(2.65)}$ & 47.14$_{(4.98)}$ & 57.87$_{(2.58)}$ & 55.98$_{(5.37)}$ \\
 & +Ours & 63.41$_{(2.88)}$ & 63.48$_{(2.11)}$ & 62.54$_{(4.24)}$ & 61.53$_{(5.50)}$ & 58.72$_{(4.03)}$ & 62.55$_{(3.26)}$ & 54.30$_{(4.15)}$ & 61.48$_{(5.04)}$ & 61.00$_{(2.90)}$ \\
 & $\Delta$ & \textcolor{gaingreen}{+1.42} & \textcolor{gaingreen}{+2.22} & \textcolor{gaingreen}{+8.75} & \textcolor{gaingreen}{+6.82} & \textcolor{gaingreen}{+9.43} & \textcolor{gaingreen}{+0.75} & \textcolor{gaingreen}{+7.16} & \textcolor{gaingreen}{+3.61} & \textcolor{gaingreen}{+5.02} \\
\midrule
\multirow{3}{*}{COAD}
 & Base  & 60.71$_{(6.18)}$ & 60.76$_{(6.86)}$ & 59.62$_{(8.47)}$ & 55.99$_{(4.25)}$ & 48.34$_{(3.31)}$ & 60.44$_{(7.28)}$ & 49.97$_{(8.26)}$ & 57.61$_{(8.23)}$ & 56.68$_{(4.63)}$ \\
 & +Ours & 62.48$_{(5.70)}$ & 62.68$_{(4.90)}$ & 64.53$_{(6.37)}$ & 60.49$_{(2.59)}$ & 58.75$_{(2.68)}$ & 61.49$_{(6.47)}$ & 58.28$_{(1.89)}$ & 65.23$_{(2.27)}$ & 61.74$_{(2.34)}$ \\
 & $\Delta$ & \textcolor{gaingreen}{+1.77} & \textcolor{gaingreen}{+1.91} & \textcolor{gaingreen}{+4.91} & \textcolor{gaingreen}{+4.49} & \textcolor{gaingreen}{+10.41} & \textcolor{gaingreen}{+1.05} & \textcolor{gaingreen}{+8.31} & \textcolor{gaingreen}{+7.61} & \textcolor{gaingreen}{+5.06} \\
\bottomrule
\end{tabular}}
\end{table}

\textbf{Molecular biomarker prediction.} \textsc{Mist} improves mean performance on every dataset, yielding an average gain of $+2.1$ points (Table~\ref{tab:biomarker_summary}), with improvements observed in 110 out of 120 configurations (Table~\ref{tab:biomarker_detailed}). MaxMIL alone achieves a $+3.0$-point gain. Because biomarker labels are defined by molecular assays and are only indirectly reflected in H\&E morphology~\cite{kather2020pan,fu2020pan}, the molecularly informed organization provided by \textsc{Mist} is particularly relevant for these tasks.

\begin{table}[t!]
\centering
\caption{\textbf{Molecular biomarker prediction.} Mean performance across biomarker tasks within each dataset, with GBMLGG additionally averaged across TCGA and EBRAINS. Results use matched 5-fold CV splits for Base and MIST (10-fold for BCNB); standard errors across folds are in parentheses.}
\label{tab:biomarker_summary}
\resizebox{\textwidth}{!}{%
\begin{tabular}{llccccccccc}
\toprule
Dataset & Status & ABMIL & CLAM & TransMIL & Trans. & ILRA & Mean & Max & DSMIL & Avg. \\
\midrule
\multirow{3}{*}{\shortstack[l]{GBMLGG\\(2 tasks)}}
 & Base  & 77.63$_{(1.72)}$ & 77.80$_{(1.47)}$ & 74.36$_{(1.69)}$ & 76.05$_{(1.39)}$ & 70.29$_{(1.83)}$ & 76.03$_{(1.47)}$ & 73.16$_{(1.71)}$ & 76.25$_{(2.41)}$ & 75.20$_{(2.35)}$ \\
 & +Ours & 79.28$_{(1.73)}$ & 79.30$_{(1.50)}$ & 77.33$_{(1.83)}$ & 78.12$_{(1.50)}$ & 73.26$_{(1.62)}$ & 77.30$_{(1.48)}$ & 75.49$_{(1.35)}$ & 77.91$_{(1.52)}$ & 77.25$_{(1.89)}$ \\
 & $\Delta$ & \textcolor{gaingreen}{+1.65} & \textcolor{gaingreen}{+1.49} & \textcolor{gaingreen}{+2.97} & \textcolor{gaingreen}{+2.06} & \textcolor{gaingreen}{+2.97} & \textcolor{gaingreen}{+1.27} & \textcolor{gaingreen}{+2.33} & \textcolor{gaingreen}{+1.66} & \textcolor{gaingreen}{+2.05} \\
\midrule
\multirow{3}{*}{\shortstack[l]{BCNB\\(3 tasks)}}
 & Base  & 85.12$_{(2.07)}$ & 85.49$_{(2.10)}$ & 82.46$_{(2.39)}$ & 83.36$_{(2.39)}$ & 78.99$_{(3.21)}$ & 81.83$_{(2.41)}$ & 79.89$_{(3.52)}$ & 84.16$_{(2.21)}$ & 82.66$_{(2.20)}$ \\
 & +Ours & 86.38$_{(1.88)}$ & 86.41$_{(1.94)}$ & 84.36$_{(2.06)}$ & 86.02$_{(1.88)}$ & 82.03$_{(2.63)}$ & 83.31$_{(2.06)}$ & 83.21$_{(2.06)}$ & 84.86$_{(2.07)}$ & 84.57$_{(1.53)}$ \\
 & $\Delta$ & \textcolor{gaingreen}{+1.26} & \textcolor{gaingreen}{+0.92} & \textcolor{gaingreen}{+1.89} & \textcolor{gaingreen}{+2.67} & \textcolor{gaingreen}{+3.04} & \textcolor{gaingreen}{+1.48} & \textcolor{gaingreen}{+3.32} & \textcolor{gaingreen}{+0.69} & \textcolor{gaingreen}{+1.91} \\
\midrule
\multirow{3}{*}{\shortstack[l]{BRCA\\(4 tasks)}}
 & Base  & 76.35$_{(2.38)}$ & 76.36$_{(2.90)}$ & 71.96$_{(2.63)}$ & 72.22$_{(2.61)}$ & 68.11$_{(1.88)}$ & 73.33$_{(2.32)}$ & 71.16$_{(2.70)}$ & 73.70$_{(2.40)}$ & 72.90$_{(2.55)}$ \\
 & +Ours & 77.16$_{(1.90)}$ & 77.21$_{(1.85)}$ & 75.25$_{(1.94)}$ & 75.64$_{(1.90)}$ & 71.42$_{(2.15)}$ & 74.32$_{(2.03)}$ & 74.11$_{(2.03)}$ & 74.83$_{(2.25)}$ & 74.99$_{(1.73)}$ \\
 & $\Delta$ & \textcolor{gaingreen}{+0.81} & \textcolor{gaingreen}{+0.84} & \textcolor{gaingreen}{+3.29} & \textcolor{gaingreen}{+3.42} & \textcolor{gaingreen}{+3.32} & \textcolor{gaingreen}{+0.99} & \textcolor{gaingreen}{+2.96} & \textcolor{gaingreen}{+1.13} & \textcolor{gaingreen}{+2.10} \\
\midrule
\multirow{3}{*}{\shortstack[l]{Lung\\(4 tasks)}}
 & Base  & 77.79$_{(2.48)}$ & 77.67$_{(2.32)}$ & 74.73$_{(2.00)}$ & 74.81$_{(1.92)}$ & 72.06$_{(1.53)}$ & 76.18$_{(1.85)}$ & 69.83$_{(3.05)}$ & 75.01$_{(2.86)}$ & 74.76$_{(2.53)}$ \\
 & +Ours & 79.29$_{(1.83)}$ & 79.44$_{(1.60)}$ & 77.68$_{(2.18)}$ & 77.97$_{(2.54)}$ & 75.26$_{(1.83)}$ & 77.16$_{(1.95)}$ & 73.00$_{(2.18)}$ & 77.24$_{(2.11)}$ & 77.13$_{(1.99)}$ \\
 & $\Delta$ & \textcolor{gaingreen}{+1.51} & \textcolor{gaingreen}{+1.77} & \textcolor{gaingreen}{+2.95} & \textcolor{gaingreen}{+3.16} & \textcolor{gaingreen}{+3.20} & \textcolor{gaingreen}{+0.98} & \textcolor{gaingreen}{+3.18} & \textcolor{gaingreen}{+2.22} & \textcolor{gaingreen}{+2.37} \\
\bottomrule
\end{tabular}}
\end{table}

\textbf{Tissue subtyping.} \textsc{Mist} improves average performance by $+3.3$ points across six tissue subtyping tasks from four datasets (Table~\ref{tab:classification_summary}), with gains observed in 59 out of 64 configurations (Table~\ref{tab:classification_detailed}); MaxMIL alone improves by $+4.5$ points. The largest gains occur on BRACS, where balanced accuracy increases by $+6.39$ points on BRACS-C and $+3.37$ points on BRACS-F. Unlike NSCLC or PANDA, several BRACS categories exhibit substantial molecular and biological overlap despite similar histological appearance, making them difficult to separate using morphology alone. In such settings, the molecularly informed organization provided by \textsc{Mist} helps emphasize subtle morphological cues that pathology foundation models alone do not prioritize. The gains are comparatively smaller on saturated cohorts where baseline H\&E performance already exceeds 90\% AUROC or weighted $\kappa$ (NSCLC and PANDA), suggesting limited remaining headroom for improvement. Further details are provided in Appendix~\ref{app:results}. With CONCH~\cite{lu2024visual} as the encoder, \textsc{Mist} also improves BRCA survival (62.50$\rightarrow$65.99), BRCA biomarkers (72.15$\rightarrow$73.77), and NSCLC (97.52$\rightarrow$97.71) on average over the eight aggregators (Appendix~\ref{app:conch}).

\begin{table}[t!]
\centering
\caption{\textbf{Tissue subtyping.} Mean performance across tissue subtyping tasks using eight MIL aggregators. NSCLC is averaged across TCGA, CPTAC, and NLST. NSCLC uses 5-fold CV; PANDA, BRACS, and EBRAINS use predefined splits. Standard errors (in parentheses) are from 1{,}000 bootstrap resamples of the test set.}
\label{tab:classification_summary}
\resizebox{\textwidth}{!}{%
\begin{tabular}{llccccccccc}
\toprule
Task & Status & ABMIL & CLAM & TransMIL & Trans. & ILRA & Mean & Max & DSMIL & Avg. \\
\midrule
\multirow{3}{*}{NSCLC (AUROC)}
 & Base  & 93.92$_{(0.75)}$ & 94.46$_{(0.74)}$ & 92.19$_{(1.32)}$ & 92.99$_{(1.38)}$ & 89.09$_{(1.44)}$ & 90.33$_{(0.82)}$ & 93.70$_{(0.80)}$ & 93.43$_{(0.93)}$ & 92.51$_{(1.76)}$ \\
 & +Ours & 94.55$_{(0.73)}$ & 94.56$_{(0.71)}$ & 93.40$_{(1.05)}$ & 93.92$_{(1.09)}$ & 91.27$_{(1.32)}$ & 91.48$_{(0.93)}$ & 94.30$_{(0.71)}$ & 94.41$_{(0.74)}$ & 93.49$_{(1.27)}$ \\
 & $\Delta$ & \textcolor{gaingreen}{+0.63} & \textcolor{gaingreen}{+0.10} & \textcolor{gaingreen}{+1.21} & \textcolor{gaingreen}{+0.93} & \textcolor{gaingreen}{+2.19} & \textcolor{gaingreen}{+1.15} & \textcolor{gaingreen}{+0.60} & \textcolor{gaingreen}{+0.98} & \textcolor{gaingreen}{+0.97} \\
\midrule
\multirow{3}{*}{PANDA (Weighted $\kappa$)}
 & Base  & 93.40$_{(0.65)}$ & 93.72$_{(0.60)}$ & 93.42$_{(0.62)}$ & 92.70$_{(0.77)}$ & 92.94$_{(0.76)}$ & 87.74$_{(1.18)}$ & 83.05$_{(1.64)}$ & 94.24$_{(0.60)}$ & 91.40$_{(3.69)}$ \\
 & +Ours & 94.15$_{(0.69)}$ & 94.15$_{(0.59)}$ & 93.92$_{(0.63)}$ & 94.19$_{(0.60)}$ & 95.47$_{(0.48)}$ & 92.17$_{(0.85)}$ & 86.29$_{(1.32)}$ & 93.99$_{(0.73)}$ & 93.04$_{(2.68)}$ \\
 & $\Delta$ & \textcolor{gaingreen}{+0.75} & \textcolor{gaingreen}{+0.43} & \textcolor{gaingreen}{+0.50} & \textcolor{gaingreen}{+1.48} & \textcolor{gaingreen}{+2.53} & \textcolor{gaingreen}{+4.43} & \textcolor{gaingreen}{+3.25} & \textcolor{lossred}{-0.24} & \textcolor{gaingreen}{+1.64} \\
\midrule
\multirow{3}{*}{BRACS-C (Bal.\ acc.)}
 & Base  & 62.59$_{(5.18)}$ & 62.41$_{(5.13)}$ & 65.99$_{(4.21)}$ & 61.82$_{(4.71)}$ & 44.70$_{(5.34)}$ & 54.89$_{(5.38)}$ & 60.28$_{(5.02)}$ & 57.61$_{(5.19)}$ & 58.79$_{(6.18)}$ \\
 & +Ours & 72.78$_{(4.79)}$ & 70.29$_{(4.90)}$ & 70.70$_{(4.97)}$ & 72.60$_{(4.67)}$ & 51.40$_{(5.14)}$ & 58.42$_{(5.49)}$ & 65.31$_{(5.48)}$ & 59.87$_{(5.14)}$ & 65.17$_{(7.36)}$ \\
 & $\Delta$ & \textcolor{gaingreen}{+10.19} & \textcolor{gaingreen}{+7.88} & \textcolor{gaingreen}{+4.71} & \textcolor{gaingreen}{+10.78} & \textcolor{gaingreen}{+6.70} & \textcolor{gaingreen}{+3.53} & \textcolor{gaingreen}{+5.03} & \textcolor{gaingreen}{+2.26} & \textcolor{gaingreen}{+6.39} \\
\midrule
\multirow{3}{*}{BRACS-F (Bal.\ acc.)}
 & Base  & 36.66$_{(4.70)}$ & 44.02$_{(4.60)}$ & 38.98$_{(4.79)}$ & 41.15$_{(5.17)}$ & 29.93$_{(4.60)}$ & 31.14$_{(4.60)}$ & 28.41$_{(4.24)}$ & 34.59$_{(5.19)}$ & 35.61$_{(5.23)}$ \\
 & +Ours & 44.35$_{(4.97)}$ & 45.22$_{(5.14)}$ & 45.58$_{(5.22)}$ & 41.90$_{(4.58)}$ & 30.25$_{(4.24)}$ & 32.20$_{(4.82)}$ & 36.02$_{(4.52)}$ & 36.36$_{(4.91)}$ & 38.98$_{(5.68)}$ \\
 & $\Delta$ & \textcolor{gaingreen}{+7.69} & \textcolor{gaingreen}{+1.20} & \textcolor{gaingreen}{+6.60} & \textcolor{gaingreen}{+0.75} & \textcolor{gaingreen}{+0.31} & \textcolor{gaingreen}{+1.06} & \textcolor{gaingreen}{+7.61} & \textcolor{gaingreen}{+1.77} & \textcolor{gaingreen}{+3.37} \\
\midrule
\multirow{3}{*}{EBRAINS-C (Bal.\ acc.)}
 & Base  & 90.77$_{(1.92)}$ & 89.86$_{(1.97)}$ & 87.97$_{(2.02)}$ & 88.11$_{(2.04)}$ & 84.74$_{(2.39)}$ & 86.97$_{(2.21)}$ & 75.91$_{(2.35)}$ & 83.24$_{(2.36)}$ & 85.95$_{(4.44)}$ \\
 & +Ours & 91.86$_{(1.84)}$ & 92.15$_{(1.79)}$ & 90.41$_{(1.94)}$ & 91.74$_{(1.68)}$ & 86.17$_{(2.25)}$ & 90.36$_{(2.02)}$ & 83.02$_{(2.26)}$ & 90.34$_{(2.04)}$ & 89.51$_{(3.02)}$ \\
 & $\Delta$ & \textcolor{gaingreen}{+1.08} & \textcolor{gaingreen}{+2.28} & \textcolor{gaingreen}{+2.45} & \textcolor{gaingreen}{+3.64} & \textcolor{gaingreen}{+1.43} & \textcolor{gaingreen}{+3.39} & \textcolor{gaingreen}{+7.12} & \textcolor{gaingreen}{+7.10} & \textcolor{gaingreen}{+3.56} \\
\midrule
\multirow{3}{*}{EBRAINS-F (Bal.\ acc.)}
 & Base  & 69.36$_{(2.20)}$ & 71.04$_{(2.19)}$ & 72.76$_{(2.00)}$ & 72.10$_{(1.96)}$ & 63.46$_{(2.21)}$ & 69.08$_{(2.20)}$ & 60.23$_{(2.28)}$ & 64.25$_{(2.20)}$ & 67.79$_{(4.28)}$ \\
 & +Ours & 74.87$_{(2.07)}$ & 74.15$_{(1.97)}$ & 73.82$_{(2.08)}$ & 75.38$_{(1.97)}$ & 68.12$_{(2.17)}$ & 73.79$_{(2.09)}$ & 63.78$_{(2.17)}$ & 68.36$_{(2.18)}$ & 71.53$_{(3.95)}$ \\
 & $\Delta$ & \textcolor{gaingreen}{+5.51} & \textcolor{gaingreen}{+3.11} & \textcolor{gaingreen}{+1.06} & \textcolor{gaingreen}{+3.28} & \textcolor{gaingreen}{+4.66} & \textcolor{gaingreen}{+4.70} & \textcolor{gaingreen}{+3.54} & \textcolor{gaingreen}{+4.11} & \textcolor{gaingreen}{+3.75} \\
\bottomrule
\end{tabular}}
\end{table}

\begin{figure}[t]
    \centering
    \includegraphics[width=0.95\linewidth]{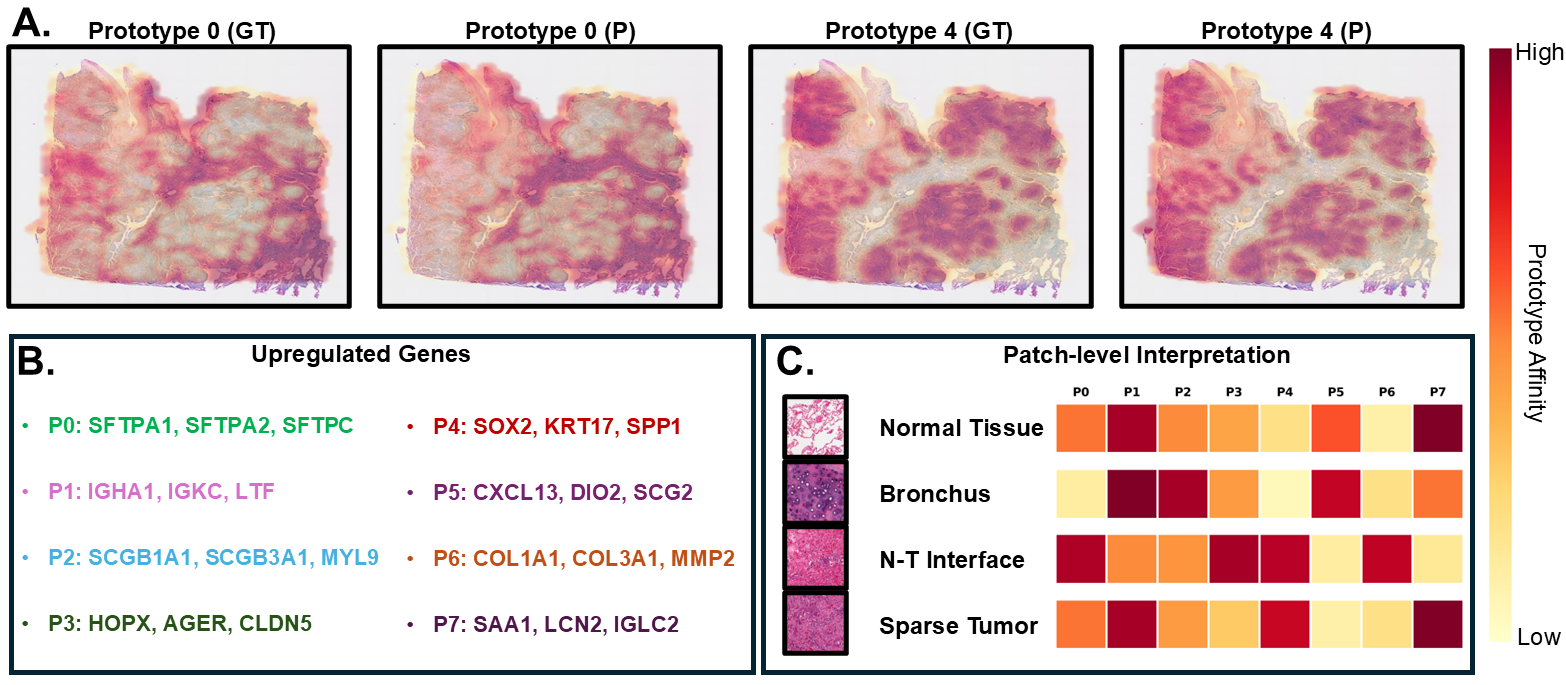}
    \caption{\textbf{Interpretability analysis of MIST.} 
    (A) Prototype affinity maps on a held-out HEST slide with paired ST. GT and P denote transcriptomics ground truth and H\&E-based prediction, respectively. 
    (B) Top upregulated genes associated with representative molecular prototypes. 
    (C) Patch-level affinity distributions across representative tissue regions annotated by board-certified pathologists, including normal tissue, bronchus, tumor-normal (N--T) interface, and sparse tumor.}
    \label{fig:interpretability}
\end{figure}

\subsection{Interpretability}
\label{sec:intrep}

\textbf{Spatial validation.}
We first evaluate whether prototypes capture spatially coherent molecular organization from H\&E alone. On a held-out HEST slide with paired ST measurements, Fig.~\ref{fig:interpretability}A compares the ground-truth spatial transcriptomic prototype maps (GT) with the corresponding model-predicted affinity maps (P) for two representative prototypes. The strong spatial agreement between GT and P indicates that MIST can infer biologically meaningful molecular distributions directly from morphology, despite receiving no ST input at inference. Notably, the predicted patterns remain spatially structured rather than diffuse, suggesting that the prototype space captures stable tissue-level molecular organization. Further details are provided in Appendix~\ref{app:more_interp}.

\textbf{Biological grounding.}
To characterize the biological meaning of each prototype, Fig.~\ref{fig:interpretability}B summarizes the top upregulated genes associated with representative molecular clusters. Several prototypes align with recognizable tissue programs and microenvironmental states. For example, P0 is enriched for surfactant-related genes (\textit{SFTPA1}, \textit{SFTPA2}, \textit{SFTPC}), consistent with alveolar epithelium, while P4 captures squamous and basal-like tumor states through markers including \textit{SOX2} and \textit{KRT17}. These two prototypes correspond closely to distinct normal and tumor-associated tissue programs visible in the slide-level and patch-level analyses. Importantly, these biological structures emerge without explicit tissue annotations during downstream training, suggesting that the prototype space is organized around intrinsic tissue biology rather than the downstream task objective alone.

\textbf{Pathological validation.}
We further examine whether the prototype affinities correspond to interpretable histopathological patterns at the local tissue level. Fig.~\ref{fig:interpretability}C shows the prototype affinity profiles of four representative patches annotated by board-certified pathologists. Each tissue state exhibits a distinct prototype signature. Normal Tissue preferentially activates alveolar-associated prototypes, whereas Sparse Tumor shows elevated tumor-associated activity. Bronchus activates airway and immune-related prototypes, while the N--T Interface co-activates alveolar, tumor, and stromal-associated prototypes, reflecting the heterogeneous composition commonly observed at invasive tumor boundaries. Importantly, these prototype signatures are not directly supervised by pathology annotations, but emerge indirectly from ST-derived molecular priors, suggesting that MIST captures clinically meaningful tissue heterogeneity through the alignment between morphology and molecular organization.

\subsection{Ablation and Hyperparameter Analysis}
\label{sec:ablation}

\textbf{Model design ablations.}
Table~\ref{tab:ablation_avg} evaluates four key design choices in MIST. Replacing gene-derived prototypes with random vectors produces the largest drop ($-2.2$), indicating that prototype structure matters. Replacing cross-modal prototypes with H\&E-only clustering, with identical architecture and capacity, still reduces performance ($-1.1$), suggesting that morphology alone captures only part of the underlying structure and that cross-modal anchoring remains important. Removing the low-rank factorization ($-1.3$) or assigning independent $\mathbf{V}_k$ to each prototype ($-1.8$) both reduce performance, supporting the design choice of learning prototype-specific modulation over a shared morphological subspace. Finally, removing either the view-level residual or the external residual leads to consistent degradation ($-1.8$), indicating that both the molecular and original pathways contribute complementary information. Using a single molecular view (72.0 on average) also underperforms the full model, supporting multi-view integration.

\textbf{Key hyperparameter.}
Table~\ref{tab:hyperparam} evaluates the number of prototypes per organ $K \in \{4,8,16,32,64\}$. MIST remains stable across most datasets, with only minor variation on EBRAINS-C/F, GBMLGG, and LUAD. BRACS shows greater sensitivity to prototype granularity, particularly on the coarse classification setting, suggesting that breast lesion subtypes may require finer molecular partitioning to resolve subtle morphological differences. Across all datasets, performance consistently peaks at small to moderate values of $K$ ($4$--$8$), while larger prototype sets provide little additional benefit. This behavior suggests that only a compact set of molecular concepts is needed to effectively organize the morphology space accessible from H\&E images. Further analysis of the computational cost and parameter overhead of MIST is provided in Appendix~\ref{app:cost}.

\textbf{Modality bottleneck.}
Table~\ref{tab:luad_orthogonal} examines whether the gains from MIST overlap with capacity-focused MIL redesigns on LUAD. Adding MAMMOTH to ABMIL improves c-index by $+0.0213$, while MIST alone yields a larger gain of $+0.0358$. Applying MIST on top of ABMIL+MAMMOTH produces an additional $+0.0323$ improvement, which remains close to the gain obtained on vanilla ABMIL. In contrast, an H\&E-only variant on top of MAMMOTH does not improve it (0.6272; Appendix~\ref{app:mammoth}), indicating that the gain is not explained by additional prototype-based capacity alone. Since both methods reorganize features at the projection stage, their gains are not expected to be strictly additive.

\begin{table}[t]
\centering
\begin{minipage}[t]{0.48\linewidth}
\centering
\caption{Ablation of key design choices in MIST, including prototype source, view transform, number of views, and residual structure. Avg. denotes the mean performance across GBMLGG, LUAD, BRACS-C/F, and EBRAINS-C/F.}
\label{tab:ablation_avg}
\small
\setlength{\tabcolsep}{4pt}
\begin{tabular}{@{}lc@{}}
\toprule
\textbf{Variant} & \textbf{Avg.} \\
\midrule
\textbf{MIST (full)} & \textbf{73.9} \\
\midrule
\multicolumn{2}{@{}l}{\textit{Prototype source (Eq.~\ref{eq:prototype})}} \\
\quad Random vectors                              & 71.7 \,{\scriptsize($-2.2$)} \\
\quad H\&E-only clustering                        & 72.8 \,{\scriptsize($-1.1$)} \\
\midrule
\multicolumn{2}{@{}l}{\textit{View transform (Eq.~\ref{eq:views})}} \\
\quad Full-rank $\mathbf{W}_k$                    & 72.6 \,{\scriptsize($-1.3$)} \\
\quad Unshared $\mathbf{V}_k$ per prototype       & 72.1 \,{\scriptsize($-1.8$)} \\
\midrule
\multicolumn{2}{@{}l}{\textit{Number of views}} \\
\quad Single view (avg.)                          & 72.0 \,{\scriptsize($-1.9$)} \\
\midrule
\multicolumn{2}{@{}l}{\textit{Residual structure}} \\
\quad No external residual (Eq.~\ref{eq:staining}) & 72.1 \,{\scriptsize($-1.8$)} \\
\quad No view residual (Eq.~\ref{eq:views})        & 72.1 \,{\scriptsize($-1.8$)} \\
\bottomrule
\end{tabular}
\end{minipage}%
\hfill
\begin{minipage}[t]{0.48\linewidth}
\centering
\caption{Sensitivity analysis over the number of molecular prototypes $K$.}
\label{tab:hyperparam}
\footnotesize
\setlength{\tabcolsep}{3pt}
\begin{tabular}{l|ccccc}
\toprule
Dataset    & $k{=}4$ & $k{=}8$ & $k{=}16$ & $k{=}32$ & $k{=}64$ \\
\midrule
BRACS-C    & 68.66 & 72.78 & 70.92 & 67.84 & 68.66 \\
BRACS-F    & 44.45 & 44.35 & 43.45 & 42.97 & 44.11 \\
EBRAINS-C  & 91.86 & 91.57 & 91.71 & 91.84 & 91.58 \\
EBRAINS-F  & 74.87 & 74.37 & 74.29 & 74.45 & 74.01 \\
GBMLGG     & 79.28 & 79.26 & 79.23 & 79.27 & 78.89 \\
LUAD       & 67.47 & 67.49 & 67.24 & 67.42 & 66.94 \\
\bottomrule
\end{tabular}

\vspace{1em}

\caption{Performance of MIST with ABMIL and ABMIL+MAMMOTH on LUAD. Results are reported as c-index.}
\label{tab:luad_orthogonal}
\small
\setlength{\tabcolsep}{6pt}
\begin{tabular}{@{}lcc@{}}
\toprule
                & w/o MIST & w/ MIST \\
\midrule
ABMIL           & 0.6111   & 0.6496  \\
ABMIL + MAMMOTH & 0.6324   & 0.6647 \\
\bottomrule
\end{tabular}
\end{minipage}
\end{table}
\label{results}

\section{Conclusion}
We introduced MIST, a plug-and-play replacement for the MIL projection layer that injects ST priors into frozen pathology foundation model pipelines while requiring only H\&E images at inference. Unlike prior ST-based approaches that adapt the encoder, MIST performs cross-modal adaptation at the projection layer, so that adopting a new pathology foundation model only requires rebuilding the prototype bank. By constructing cross-modal prototypes from paired ST and H\&E data, MIST reorganizes patch representations according to molecular programs. Across 23 downstream tasks and 8 MIL aggregators, MIST consistently improves performance, particularly on survival prediction. Gains are strongest for weaker aggregators and molecularly heterogeneous tasks, consistent with the view that the MIL projection layer is limited not only by capacity but also by the lack of molecular supervision. Interpretability analyses further show that learned prototype affinities capture spatially coherent tissue programs directly from H\&E images.
\label{conclusion}

\begin{ack}
This research is supported by the National Medical Research Council (NMRC) Healthy and Meaningful Longevity -- Cognition Grant (NICCOG2024-0028) and the AI for Public Health Program in Saw Swee Hock School of Public Health, National University of Singapore. This work was also supported by the Dame Julia Higgins Postdoc Collaborative Research Fund, the NUS-Guangzhou Research Translation and Innovation Institute Scholarship, the National Natural Science Foundation of China (Grant No.~62602189), and the China Postdoctoral Science Foundation (Grant Nos.~2026M791622, 2026T019HN, and GZB20260396). The authors declare no competing interests.
\end{ack}

\clearpage
\bibliographystyle{unsrt}
\bibliography{neurips_2026}

@article{coudray2018classification,
  title={Classification and mutation prediction from non--small cell lung cancer histopathology images using deep learning},
  author={Coudray, Nicolas and Ocampo, Paolo Santiago and Sakellaropoulos, Theodore and Narula, Navneet and Snuderl, Matija and Feny{\"o}, David and Moreira, Andre L and Razavian, Narges and Tsirigos, Aristotelis},
  journal={Nature medicine},
  volume={24},
  number={10},
  pages={1559--1567},
  year={2018},
  publisher={Nature Publishing Group US New York}
}

@article{kather2019deep,
  title={Deep learning can predict microsatellite instability directly from histology in gastrointestinal cancer},
  author={Kather, Jakob Nikolas and Pearson, Alexander T and Halama, Niels and J{\"a}ger, Dirk and Krause, Jeremias and Loosen, Sven H and Marx, Alexander and Boor, Peter and Tacke, Frank and Neumann, Ulf Peter and others},
  journal={Nature medicine},
  volume={25},
  number={7},
  pages={1054--1056},
  year={2019},
  publisher={Nature Publishing Group US New York}
}

@article{campanella2019clinical,
  title={Clinical-grade computational pathology using weakly supervised deep learning on whole slide images},
  author={Campanella, Gabriele and Hanna, Matthew G and Geneslaw, Luke and Miraflor, Allen and Werneck Krauss Silva, Vitor and Busam, Klaus J and Brogi, Edi and Reuter, Victor E and Klimstra, David S and Fuchs, Thomas J},
  journal={Nature medicine},
  volume={25},
  number={8},
  pages={1301--1309},
  year={2019},
  publisher={Nature Publishing Group US New York}
}

@article{hoadley2018cell,
  title={Cell-of-origin patterns dominate the molecular classification of 10,000 tumors from 33 types of cancer},
  author={Hoadley, Katherine A and Yau, Christina and Hinoue, Toshinori and Wolf, Denise M and Lazar, Alexander J and Drill, Esther and Shen, Ronglai and Taylor, Alison M and Cherniack, Andrew D and Thorsson, V{\'e}steinn and others},
  journal={Cell},
  volume={173},
  number={2},
  pages={291--304},
  year={2018},
  publisher={Elsevier}
}

@incollection{magaki2018introduction,
  title={An introduction to the performance of immunohistochemistry},
  author={Magaki, Shino and Hojat, Seyed A and Wei, Bowen and So, Alexandra and Yong, William H},
  booktitle={Biobanking: methods and protocols},
  pages={289--298},
  year={2018},
  publisher={Springer}
}

@article{xiang2026multimodal,
  title={A Multimodal Foundation Model of Spatial Transcriptomics and Histology for Biological Discovery and Clinical Prediction},
  author={Xiang, Jinxi and Hou, Siyu and Li, Yuchen and Quinton, Ryan and Zhang, Xiaoming and Eweje, Feyisope and Luo, Xiangde and Chen, Yijiang and Li, Zhe and Bergstrom, Colin and others},
  journal={arXiv preprint arXiv:2604.03630},
  year={2026}
}

@article{vorontsov2024foundation,
  title={A foundation model for clinical-grade computational pathology and rare cancers detection},
  author={Vorontsov, Eugene and Bozkurt, Alican and Casson, Adam and Shaikovski, George and Zelechowski, Michal and Severson, Kristen and Zimmermann, Eric and Hall, James and Tenenholtz, Neil and Fusi, Nicolo and others},
  journal={Nature medicine},
  volume={30},
  number={10},
  pages={2924--2935},
  year={2024},
  publisher={Nature Publishing Group US New York}
}

@article{xu2024whole,
  title={A whole-slide foundation model for digital pathology from real-world data},
  author={Xu, Hanwen and Usuyama, Naoto and Bagga, Jaspreet and Zhang, Sheng and Rao, Rajesh and Naumann, Tristan and Wong, Cliff and Gero, Zelalem and Gonz{\'a}lez, Javier and Gu, Yu and others},
  journal={Nature},
  volume={630},
  number={8015},
  pages={181--188},
  year={2024},
  publisher={Nature Publishing Group UK London}
}

@inproceedings{chen2022scaling,
  title={Scaling vision transformers to gigapixel images via hierarchical self-supervised learning},
  author={Chen, Richard J and Chen, Chengkuan and Li, Yicong and Chen, Tiffany Y and Trister, Andrew D and Krishnan, Rahul G and Mahmood, Faisal},
  booktitle={Proceedings of the IEEE/CVF conference on computer vision and pattern recognition},
  pages={16144--16155},
  year={2022}
}

@article{lu2024visual,
  title={A visual-language foundation model for computational pathology},
  author={Lu, Ming Y and Chen, Bowen and Williamson, Drew FK and Chen, Richard J and Liang, Ivy and Ding, Tong and Jaume, Guillaume and Odintsov, Igor and Le, Long Phi and Gerber, Georg and others},
  journal={Nature medicine},
  volume={30},
  number={3},
  pages={863--874},
  year={2024},
  publisher={Nature Publishing Group US New York}
}

@article{ding2025multimodal,
  title={A multimodal whole-slide foundation model for pathology},
  author={Ding, Tong and Wagner, Sophia J and Song, Andrew H and Chen, Richard J and Lu, Ming Y and Zhang, Andrew and Vaidya, Anurag J and Jaume, Guillaume and Shaban, Muhammad and Kim, Ahrong and others},
  journal={Nature medicine},
  pages={1--13},
  year={2025},
  publisher={Nature Publishing Group US New York}
}

@inproceedings{tang2024feature,
  title={Feature re-embedding: Towards foundation model-level performance in computational pathology},
  author={Tang, Wenhao and Zhou, Fengtao and Huang, Sheng and Zhu, Xiang and Zhang, Yi and Liu, Bo},
  booktitle={Proceedings of the IEEE/CVF conference on computer vision and pattern recognition},
  pages={11343--11352},
  year={2024}
}

@article{guo2025context,
  title={Context matters: Query-aware dynamic long sequence modeling of gigapixel images},
  author={Guo, Zhengrui and Sun, Qichen and Ma, Jiabo and Feng, Lishuang and Wang, Jinzhuo and Chen, Hao},
  journal={arXiv preprint arXiv:2501.18984},
  year={2025}
}

@article{chen2022pan,
  title={Pan-cancer integrative histology-genomic analysis via multimodal deep learning},
  author={Chen, Richard J and Lu, Ming Y and Williamson, Drew FK and Chen, Tiffany Y and Lipkova, Jana and Noor, Zahra and Shaban, Muhammad and Shady, Maha and Williams, Mane and Joo, Bumjin and others},
  journal={Cancer cell},
  volume={40},
  number={8},
  pages={865--878},
  year={2022},
  publisher={Elsevier}
}

@article{lee2026mint,
  title={MINT: Molecularly Informed Training with Spatial Transcriptomics Supervision for Pathology Foundation Models},
  author={Lee, Minsoo and Kim, Jonghyun and Yun, Juseung and Yu, Sunwoo and Jang, Jongseong},
  journal={arXiv preprint arXiv:2603.07895},
  year={2026}
}

@article{redekop2025spade,
  title={Spade: Spatial transcriptomics and pathology alignment using a mixture of data experts for an expressive latent space},
  author={Redekop, Ekaterina and Pleasure, Mara and Wang, Zichen and Flores, Kimberly and Sisk, Anthony and Speier, William and Arnold, Corey W},
  journal={arXiv preprint arXiv:2506.21857},
  year={2025}
}

@inproceedings{xing2025dpsurv,
  author    = {Xing, Yucheng and Huang, Ling and Ma, Jingying and Hong, Ruping and Qiu, Jiangdong and Liu, Pei and He, Kai and Fu, Huazhu and Feng, Mengling},
  title     = {{DPsurv}: Dual-Prototype Evidential Fusion for Uncertainty-Aware and Interpretable Whole Slide Image Survival Prediction},
  booktitle = {International Conference on Machine Learning (ICML)},
  pages     = {140060--140084},
  year      = {2026}
}

@inproceedings{xingemms,
  author    = {Xing, Yucheng and Mo, Hailan and Wang, Zi and Huang, Ling and Feng, Mengling},
  title     = {Evidential Fusion Network for Multimodal Survival Prediction under Missing Modalities},
  booktitle = {International Conference on Medical Image Computing and Computer-Assisted Intervention (MICCAI)},
  pages     = {238--248},
  year      = {2026}
}

@inproceedings{xingsaefs,
  author    = {Xing, Yucheng and Huang, Ling and Liu, Pei and Ma, Jingying and Xu, Jiaxing and He, Kai and Feng, Mengling},
  title     = {Semantic-Anchored Evidential Fusion for Domain-Robust Whole-Slide Survival Analysis},
  booktitle = {European Conference on Computer Vision (ECCV)},
  pages     = {150--169},
  year      = {2026}
}

@article{masscope,
  author  = {Ma, Jingying and Wu, Feng and Xing, Yucheng and Lin, Qika and Liu, Tianyu and Liu, Chenyu and Jia, Ziyu and Feng, Mengling},
  title   = {{SCOPE}: Structured Prototype-Guided Adaptation for {EEG} Foundation Models with Limited Labels},
  journal = {arXiv preprint arXiv:2602.17251},
  year    = {2026}
}

@article{huang2025esurvfusion,
  title={EsurvFusion: An evidential multimodal survival fusion model based on Epistemic random fuzzy sets},
  author={Huang, Ling and Xing, Yucheng and Lin, Qika and Duan, Jinming and Ruan, Su and Feng, Mengling},
  journal={IEEE Transactions on Fuzzy Systems},
  year={2025},
  publisher={IEEE}
}

@inproceedings{huang2024evidential,
  title={An evidential time-to-event prediction model based on Gaussian random fuzzy numbers},
  author={Huang, Ling and Xing, Yucheng and Denoeux, Thierry and Feng, Mengling},
  booktitle={International Conference on Belief Functions},
  pages={49--57},
  year={2024},
  organization={Springer}
}

@article{huang2025evidential,
  title={Evidential time-to-event prediction with calibrated uncertainty quantification},
  author={Huang, Ling and Xing, Yucheng and Mishra, Swapnil and Den{\oe}ux, Thierry and Feng, Mengling},
  journal={International Journal of Approximate Reasoning},
  volume={181},
  pages={109403},
  year={2025},
  publisher={Elsevier}
}

@inproceedings{song2024morphological,
    title={Morphological Prototyping for Unsupervised Slide Representation Learning in Computational Pathology},
    author={Song, Andrew H and Chen, Richard J and Ding, Tong and Williamson, Drew FK and Jaume, Guillaume and Mahmood, Faisal},
    booktitle={Proceedings of the IEEE/CVF Conference on Computer Vision and Pattern Recognition},
    year={2024},
}

@article{vu2023handcrafted,
  title={Handcrafted Histological Transformer (H2T): Unsupervised representation of whole slide images},
  author={Vu, Quoc Dang and Rajpoot, Kashif and Raza, Shan E Ahmed and Rajpoot, Nasir},
  journal={Medical image analysis},
  volume={85},
  pages={102743},
  year={2023},
  publisher={Elsevier}
}

@article{song2024multimodal,
  title={Multimodal prototyping for cancer survival prediction},
  author={Song, Andrew H and Chen, Richard J and Jaume, Guillaume and Vaidya, Anurag J and Baras, Alexander S and Mahmood, Faisal},
  journal={arXiv preprint arXiv:2407.00224},
  year={2024}
}

@article{xiang2025vision,
  title={A vision--language foundation model for precision oncology},
  author={Xiang, Jinxi and Wang, Xiyue and Zhang, Xiaoming and Xi, Yinghua and Eweje, Feyisope and Chen, Yijiang and Li, Yuchen and Bergstrom, Colin and Gopaulchan, Matthew and Kim, Ted and others},
  journal={Nature},
  volume={638},
  number={8051},
  pages={769--778},
  year={2025},
  publisher={Nature Publishing Group UK London}
}

@inproceedings{shao2026mammoth,
  title={Mixture of Mini Experts: Overcoming the Linear Layer Bottleneck in Multiple Instance Learning},
  author={Shao, Daniel and Runevic, Joel and Chen, Richard J. and Williamson, Drew F. K. and Kim, Ahrong and Song, Andrew H. and Mahmood, Faisal},
  booktitle={International Conference on Learning Representations (ICLR)},
  year={2026},
  url={https://openreview.net/forum?id=S5Io33pc78}
}

@inproceedings{jaume2024hest,
    author = {Guillaume Jaume and Paul Doucet and Andrew H. Song and Ming Y. Lu and Cristina Almagro-Perez and Sophia J. Wagner and Anurag J. Vaidya and Richard J. Chen and Drew F. K. Williamson and Ahrong Kim and Faisal Mahmood},
    title = {HEST-1k: A Dataset for Spatial Transcriptomics and Histology Image Analysis},
    booktitle = {Advances in Neural Information Processing Systems},
    year = {2024},
    month = dec,
}

@article{chen2024uni,
  title={Towards a General-Purpose Foundation Model for Computational Pathology},
  author={Chen, Richard J and Ding, Tong and Lu, Ming Y and Williamson, Drew FK and Jaume, Guillaume and Chen, Bowen and Zhang, Andrew and Shao, Daniel and Song, Andrew H and Shaban, Muhammad and others},
  journal={Nature Medicine},
  publisher={Nature Publishing Group},
  year={2024}
}

@article{hemker2026seal,
    author = {Konstantin Hemker and Andrew H. Song and Cristina Almagro-Perez and Guillaume Jaume and Sophie J. Wagner and Anurag Vaidya and Nikola Simidjievski and Mateja Jamnik and Faisal Mahmood}, 
    title = {Towards Spatial Transcriptomics-driven Pathology Foundation Models},
    journal = {arXiv preprint arXiv:2602.14177}, 
    year = {2026}
}

@article{stuart2019comprehensive,
  title={Comprehensive integration of single-cell data},
  author={Stuart, Tim and Butler, Andrew and Hoffman, Paul and Hafemeister, Christoph and Papalexi, Efthymia and Mauck, William M and Hao, Yuhan and Stoeckius, Marlon and Smibert, Peter and Satija, Rahul},
  journal={cell},
  volume={177},
  number={7},
  pages={1888--1902},
  year={2019},
  publisher={Elsevier}
}

@article{roetzer2022digital,
  title={The digital brain tumour atlas, an open histopathology resource},
  author={Roetzer-Pejrimovsky, Thomas and Moser, Anna-Christina and Atli, Baran and Vogel, Clemens Christian and Mercea, Petra A and Prihoda, Romana and Gelpi, Ellen and Haberler, Christine and H{\"o}ftberger, Romana and Hainfellner, Johannes A and others},
  journal={Scientific Data},
  volume={9},
  number={1},
  pages={55},
  year={2022},
  publisher={Nature Publishing Group UK London}
}

@article{campbell2016distinct,
  title={Distinct patterns of somatic genome alterations in lung adenocarcinomas and squamous cell carcinomas},
  author={Campbell, Joshua D and Alexandrov, Anton and Kim, Jaegil and Wala, Jeremiah and Berger, Alice H and Pedamallu, Chandra Sekhar and Shukla, Sachet A and Guo, Guangwu and Brooks, Angela N and Murray, Bradley A and others},
  journal={Nature genetics},
  volume={48},
  number={6},
  pages={607--616},
  year={2016},
  publisher={Nature Publishing Group US New York}
}

@article{bulten2022artificial,
  title={Artificial intelligence for diagnosis and Gleason grading of prostate cancer: the PANDA challenge},
  author={Bulten, Wouter and Kartasalo, Kimmo and Chen, Po-Hsuan Cameron and Str{\"o}m, Peter and Pinckaers, Hans and Nagpal, Kunal and Cai, Yuannan and Steiner, David F and Van Boven, Hester and Vink, Robert and others},
  journal={Nature medicine},
  volume={28},
  number={1},
  pages={154--163},
  year={2022},
  publisher={Nature Publishing Group US New York}
}

@article{brancati2022bracs,
  title={Bracs: A dataset for breast carcinoma subtyping in h\&e histology images},
  author={Brancati, Nadia and Anniciello, Anna Maria and Pati, Pushpak and Riccio, Daniel and Scognamiglio, Giosu{\`e} and Jaume, Guillaume and De Pietro, Giuseppe and Di Bonito, Maurizio and Foncubierta, Antonio and Botti, Gerardo and others},
  journal={Database},
  volume={2022},
  pages={baac093},
  year={2022},
  publisher={Oxford University Press UK}
}

@article{xu2021predicting,
  title={Predicting axillary lymph node metastasis in early breast cancer using deep learning on primary tumor biopsy slides},
  author={Xu, Feng and Zhu, Chuang and Tang, Wenqi and Wang, Ying and Zhang, Yu and Li, Jie and Jiang, Hongchuan and Shi, Zhongyue and Liu, Jun and Jin, Mulan},
  journal={Frontiers in oncology},
  volume={11},
  pages={759007},
  year={2021},
  publisher={Frontiers}
}

@article{weinstein2013cancer,
  title={The cancer genome atlas pan-cancer analysis project},
  author={Weinstein, John N and Collisson, Eric A and Mills, Gordon B and Shaw, Kenna R and Ozenberger, Brad A and Ellrott, Kyle and Shmulevich, Ilya and Sander, Chris and Stuart, Joshua M},
  journal={Nature genetics},
  volume={45},
  number={10},
  pages={1113--1120},
  year={2013},
  publisher={Nature Publishing Group}
}

@inproceedings{ilse2018attention,
  title={Attention-based deep multiple instance learning},
  author={Ilse, Maximilian and Tomczak, Jakub and Welling, Max},
  booktitle={International conference on machine learning},
  pages={2127--2136},
  year={2018},
  organization={PMLR}
}

@article{lu2021data,
  title={Data-efficient and weakly supervised computational pathology on whole-slide images},
  author={Lu, Ming Y and Williamson, Drew FK and Chen, Tiffany Y and Chen, Richard J and Barbieri, Matteo and Mahmood, Faisal},
  journal={Nature biomedical engineering},
  volume={5},
  number={6},
  pages={555--570},
  year={2021},
  publisher={Nature Publishing Group UK London}
}

@inproceedings{li2021dual,
  title={Dual-stream multiple instance learning network for whole slide image classification with self-supervised contrastive learning},
  author={Li, Bin and Li, Yin and Eliceiri, Kevin W},
  booktitle={Proceedings of the IEEE/CVF conference on computer vision and pattern recognition},
  pages={14318--14328},
  year={2021}
}

@article{shao2021transmil,
  title={Transmil: Transformer based correlated multiple instance learning for whole slide image classification},
  author={Shao, Zhuchen and Bian, Hao and Chen, Yang and Wang, Yifeng and Zhang, Jian and Ji, Xiangyang and others},
  journal={Advances in neural information processing systems},
  volume={34},
  pages={2136--2147},
  year={2021}
}

@inproceedings{xiang2023exploring,
  title={Exploring low-rank property in multiple instance learning for whole slide image classification},
  author={Xiang, Jinxi and Zhang, Jun},
  booktitle={The Eleventh International Conference on Learning Representations},
  year={2023}
}

@article{wagner2023transformer,
  title={Transformer-based biomarker prediction from colorectal cancer histology: A large-scale multicentric study},
  author={Wagner, Sophia J and Reisenb{\"u}chler, Daniel and West, Nicholas P and Niehues, Jan Moritz and Zhu, Jiefu and Foersch, Sebastian and Veldhuizen, Gregory Patrick and Quirke, Philip and Grabsch, Heike I and van den Brandt, Piet A and others},
  journal={Cancer cell},
  volume={41},
  number={9},
  pages={1650--1661},
  year={2023},
  publisher={Elsevier}
}

@article{vaswani2017attention,
  title={Attention is all you need},
  author={Vaswani, Ashish and Shazeer, Noam and Parmar, Niki and Uszkoreit, Jakob and Jones, Llion and Gomez, Aidan N and Kaiser, {\L}ukasz and Polosukhin, Illia},
  journal={Advances in neural information processing systems},
  volume={30},
  year={2017}
}

@article{kather2020pan,
  title={Pan-cancer image-based detection of clinically actionable genetic alterations},
  author={Kather, Jakob Nikolas and Heij, Lara R and Grabsch, Heike I and Loeffler, Chiara and Echle, Amelie and Muti, Hannah Sophie and Krause, Jeremias and Niehues, Jan M and Sommer, Kai AJ and Bankhead, Peter and others},
  journal={Nature cancer},
  volume={1},
  number={8},
  pages={789--799},
  year={2020},
  publisher={Nature Publishing Group US New York}
}

@article{fu2020pan,
  title={Pan-cancer computational histopathology reveals mutations, tumor composition and prognosis},
  author={Fu, Yu and Jung, Alexander W and Torne, Ramon Vi{\~n}as and Gonzalez, Santiago and V{\"o}hringer, Harald and Shmatko, Artem and Yates, Lucy R and Jimenez-Linan, Mercedes and Moore, Luiza and Gerstung, Moritz},
  journal={Nature cancer},
  volume={1},
  number={8},
  pages={800--810},
  year={2020},
  publisher={Nature Publishing Group US New York}
}


\clearpage
\appendix

\section*{Appendix}

\section{Molecular Prototype Construction}
\label{app:prototype_construction}
We derive $K$ molecular prototypes offline, each capturing a distinct gene expression program while living in the morphological feature space so it can directly interact with patch embeddings at training time. We use the HEST-1k dataset~\cite{jaume2024hest}, which provides 1{,}276 spatially-resolved transcriptomic profiles paired with H\&E whole-slide images across 26 organ types and approximately 940{,}000 human Visium spots.
\paragraph{Data curation.}
We apply SEAL-style filtering~\cite{hemker2026seal} to select high-quality samples: (1)Homo sapiens only; (2)~Visium or Visium~HD (spot diameter $\leq 55\,\mu$m), excluding legacy 100\,$\mu$m ST arrays; (3)~14 organ types (Breast, Bowel, Skin, Kidney, Heart, Prostate, Brain, Lung, Uterus, Liver, Bladder, Pancreas, Cervix, Ovary); (4)~$\geq 5{,}000$ detected genes per sample.
\paragraph{Patch features.}
For each sample, patches at $0.5\,\mu$m/px ($20\times$) are tessellated from the H\&E slide and encoded by a frozen pathology foundation model (UNI2-h~\cite{chen2024uni}, $D{=}1536$). Each patch is matched to its overlapping Visium spot by coordinate alignment, producing paired samples $\{(\mathbf{x}_j, \mathbf{g}_j)\}_{j=1}^{M}$.
\paragraph{Gene expression preprocessing.}
After quality control (removing empty spots and genes detected in ${<}10\%$ of spots), counts are normalized to a target sum of $10^4$ per spot and transformed via $\log(1{+}x)$. We select the top 2{,}000 highly variable genes (HVGs) using the Seurat~v3 variance-stabilizing method~\cite{stuart2019comprehensive} on the raw count layer. To mitigate Visium sparsity and noise, we apply spatial smoothing within each tissue section independently to prevent leakage across sections. For each spot $j$, we average its HVG vector with its $n{=}6$ nearest neighbors by Euclidean distance on spatial coordinates (approximating the Visium hexagonal grid):
\begin{equation}
    \bar{\mathbf{g}}_j = \frac{1}{|\mathcal{N}_n(j)|+1}\Big(\mathbf{g}_j + \sum_{j' \in \mathcal{N}_n(j)} \mathbf{g}_{j'}\Big).
    \label{eq:smoothing}
\end{equation}
\paragraph{Molecular clustering.}
We run MiniBatch K-means on the smoothed vectors $\{\bar{\mathbf{g}}_j\}_{j=1}^{M}$ pooled across all samples, assigning each spot a cluster label $c_j \in \{1, \ldots, K\}$. Each cluster corresponds to a distinct molecular program, e.g., hepatocyte metabolism (SERPINA1$^+$), B-cell immune response (IGKC$^+$), glial identity (GFAP$^+$), or epithelial differentiation (KRT19$^+$). For each organ, we scan $K \in \{4, 8, 16, 32, 64\}$ and select the value with the highest silhouette score on a random subset of 50{,}000 spots (e.g., $K{=}8$ for breast and $K{=}4$ for brain).
\paragraph{Prototype computation.}
For each cluster $k$, the molecular prototype is the centroid of its patch features):
\begin{equation}
    \boldsymbol{\mu}_k = \frac{1}{|\mathcal{C}_k|} \sum_{j \in \mathcal{C}_k} \mathbf{x}_j \in \mathbb{R}^{D}, \qquad \mathcal{C}_k = \{j : c_j = k\}.
    \label{eq:prototype_appendix}
\end{equation}
This cross-modal design is central: clustering is performed in gene expression space to discover molecular states, but the prototypes live in morphological feature space so they can directly condition patch embeddings at training time. Each $\boldsymbol{\mu}_k$ thus encodes the average morphological appearance of a specific molecular program, serving as a \emph{molecular view}. The full set $\boldsymbol{\Pi} = \{\boldsymbol{\mu}_k\}_{k=1}^{K} \in \mathbb{R}^{K \times D}$ is stored offline and used to initialize MIST at the start of training.

\section{MIST Can Increase Separation Between Morphologically Similar but Molecularly Distinct Patches}
\label{app:theory}

\paragraph{Setup.}
Let $\mathbf{h}_i, \mathbf{h}_j \in \mathbb{R}^d$ be two patch embeddings 
that are morphologically similar but molecularly distinct:
\begin{equation}
    \mathbf{h}_i - \mathbf{h}_j = \epsilon\mathbf{v}, 
    \quad \|\mathbf{v}\| = 1, \quad \epsilon \ll 1,
    \label{eq:morph-similar}
\end{equation}
and patch $i$ is associated with prototype $A$ while patch $j$ is 
associated with prototype $B$, i.e., $\alpha_A(\mathbf{h}_i) > 
\alpha_A(\mathbf{h}_j)$. Denote $\Delta\alpha_k = \alpha_k(\mathbf{h}_i) 
- \alpha_k(\mathbf{h}_j)$, with $\Delta\alpha_A > 0$ and 
$\Delta\alpha_B < 0$.

\paragraph{Molecular baseline.}
Define the \emph{molecular baseline} as the MIST output when the 
affinity is uniform across prototypes, i.e., 
$\alpha_k \equiv 1/K$ for all $k$. In this case the prototype-conditioned 
term reduces to a fixed transform $\bar{\mathbf{W}} = 
\frac{1}{K}\sum_k \mathbf{U}_k \mathbf{V}$, and the output distance is
\begin{equation}
    \|f_\mathrm{base}(\mathbf{h}_i) - f_\mathrm{base}(\mathbf{h}_j)\|
    = \|(2\mathbf{I} + \bar{\mathbf{W}})\epsilon\mathbf{v}\|.
    \label{eq:baseline-dist}
\end{equation}
This baseline captures the separation that MIST achieves from 
morphology alone, with no molecular routing.

\paragraph{MIST output distance.}
The full MIST output difference (to first order in $\epsilon$) is
\begin{equation}
    \tilde{\mathbf{h}}_i - \tilde{\mathbf{h}}_j
    = (2\mathbf{I} + \bar{\mathbf{W}})\epsilon\mathbf{v}
    + \underbrace{\Bigl(\sum_k \Delta\alpha_k\, 
      \mathbf{U}_k\Bigr)\mathbf{V}\bar{\mathbf{h}}}_{\text{molecular routing term}},
    \label{eq:mist-diff}
\end{equation}
where $\bar{\mathbf{h}} = \frac{1}{2}(\mathbf{h}_i + \mathbf{h}_j)$ 
and higher-order terms in $\epsilon$ are absorbed into 
$o(\epsilon)$.

\paragraph{Proposition.}
\textit{There exists a parameter configuration 
$\{\mathbf{U}_k\}, \mathbf{V}$ such that}
\begin{equation}
    \|\tilde{\mathbf{h}}_i - \tilde{\mathbf{h}}_j\| 
    > \|f_\mathrm{base}(\mathbf{h}_i) - f_\mathrm{base}(\mathbf{h}_j)\|.
    \label{eq:separation}
\end{equation}

\paragraph{Proof.}
Let $\mathbf{n} = (2\mathbf{I} + \bar{\mathbf{W}})\mathbf{v}$ denote 
the baseline direction (normalized: $\hat{\mathbf{n}} = 
\mathbf{n}/\|\mathbf{n}\|$). Construct parameters as follows:
\begin{equation}
    \mathbf{V} = \hat{\mathbf{h}}^{\top}, 
    \qquad 
    \mathbf{U}_A - \mathbf{U}_B = c\,\hat{\mathbf{n}},
    \quad c > 0,
    \label{eq:construction}
\end{equation}
where $\hat{\mathbf{h}} = \bar{\mathbf{h}}/\|\bar{\mathbf{h}}\|$. 
Under this construction, the molecular routing term becomes
\begin{equation}
    \Bigl(\sum_k \Delta\alpha_k\,\mathbf{U}_k\Bigr)\mathbf{V}\bar{\mathbf{h}}
    = \Delta\alpha_A\,(\mathbf{U}_A - \mathbf{U}_B)\,\|\bar{\mathbf{h}}\|
    = c\,\Delta\alpha_A\,\|\bar{\mathbf{h}}\|\,\hat{\mathbf{n}},
\end{equation}
using $\sum_k \Delta\alpha_k = 0$ (from $\ell_1$ normalization). 
Since $\Delta\alpha_A > 0$, $c > 0$, and $\|\bar{\mathbf{h}}\| > 0$, 
this term is strictly positive along $\hat{\mathbf{n}}$, i.e., 
aligned with the baseline direction. Therefore,
\begin{equation}
    \|\tilde{\mathbf{h}}_i - \tilde{\mathbf{h}}_j\|
    = \|\epsilon\mathbf{n} + c\,\Delta\alpha_A\,\|\bar{\mathbf{h}}\|\,\hat{\mathbf{n}}\|
    = \epsilon\|\mathbf{n}\| + c\,\Delta\alpha_A\,\|\bar{\mathbf{h}}\|
    > \epsilon\|\mathbf{n}\|
    = \|f_\mathrm{base}(\mathbf{h}_i) - f_\mathrm{base}(\mathbf{h}_j)\|,
\end{equation}
which establishes Eq.~\eqref{eq:separation}.

\paragraph{Remark.}
The construction in Eq.~\eqref{eq:construction} is not unique; any 
$\mathbf{U}_A - \mathbf{U}_B$ with a positive projection onto 
$\hat{\mathbf{n}}$ suffices. The key condition is 
$\sum_k \Delta\alpha_k = 0$, which holds by design from the $\ell_1$ 
normalization of affinities, and $\Delta\alpha_A > 0$, which holds 
whenever the prototype affinities correctly reflect the molecular 
identity of the patch. Since affinities are computed from H\&E features,
this is an assumption on the learned prototypes rather than a guarantee.
Whether MIL training discovers such a
configuration is an empirical question; Section~\ref{sec:intrep} 
provides evidence that it does.

\section{Detailed Results}
\label{app:results}

This section summarizes the complete downstream evaluation results across survival prediction, molecular biomarker prediction, and tissue subtyping. In total, we evaluate 256 model-task configurations spanning multiple datasets, external validation cohorts, and eight MIL aggregators. MIST improves 240 configurations, with only 16 configurations showing negative changes relative to the standard projection layer.The detailed tables below show that the gains are broadly consistent across endpoint types, aggregators, and external evaluation settings (Tables~\ref{tab:biomarker_detailed}, \ref{tab:classification_detailed}, and \ref{tab:survival_detailed}).

\begin{table}[t]
\centering
\caption{\textbf{Survival prediction (detailed breakdown).} Concordance index (\%) across individual cohorts and external validation datasets using eight MIL aggregators. CPTAC and NLST are evaluated as external cohorts. Matched 5-fold CV splits for Base and MIST; standard errors across folds are in parentheses. Avg. denotes the cross-aggregator mean performance.}
\label{tab:survival_detailed}
\resizebox{\textwidth}{!}{%
\begin{tabular}{llccccccccc}
\toprule
Cohort & Status & ABMIL & CLAM & TransMIL & Trans. & ILRA & Mean & Max & DSMIL & Avg. \\
\midrule
\multirow{3}{*}{BRCA}
 & Base  & 60.58$_{(6.44)}$ & 60.79$_{(6.33)}$ & 62.66$_{(9.88)}$ & 63.12$_{(8.92)}$ & 57.47$_{(6.20)}$ & 65.77$_{(7.26)}$ & 54.95$_{(3.82)}$ & 60.99$_{(4.92)}$ & 60.79$_{(3.14)}$ \\
 & +Ours & 69.34$_{(3.87)}$ & 69.70$_{(3.65)}$ & 67.56$_{(7.98)}$ & 70.28$_{(8.02)}$ & 65.31$_{(6.37)}$ & 70.96$_{(1.54)}$ & 58.77$_{(9.11)}$ & 68.12$_{(2.51)}$ & 67.50$_{(3.70)}$ \\
 & $\Delta$ & \textcolor{gaingreen}{+8.76} & \textcolor{gaingreen}{+8.91} & \textcolor{gaingreen}{+4.90} & \textcolor{gaingreen}{+7.16} & \textcolor{gaingreen}{+7.84} & \textcolor{gaingreen}{+5.19} & \textcolor{gaingreen}{+3.81} & \textcolor{gaingreen}{+7.13} & \textcolor{gaingreen}{+6.71} \\
\midrule
\multirow{3}{*}{LUAD (TCGA)}
 & Base  & 61.11$_{(11.26)}$ & 61.00$_{(11.39)}$ & 63.45$_{(7.33)}$ & 58.80$_{(6.42)}$ & 47.53$_{(1.26)}$ & 61.40$_{(10.83)}$ & 52.26$_{(7.68)}$ & 58.76$_{(8.76)}$ & 58.04$_{(5.05)}$ \\
 & +Ours & 64.96$_{(10.77)}$ & 64.80$_{(10.85)}$ & 63.85$_{(15.56)}$ & 60.27$_{(9.55)}$ & 47.88$_{(2.83)}$ & 63.37$_{(11.57)}$ & 55.05$_{(4.89)}$ & 62.75$_{(9.87)}$ & 60.37$_{(5.61)}$ \\
 & $\Delta$ & \textcolor{gaingreen}{+3.85} & \textcolor{gaingreen}{+3.81} & \textcolor{gaingreen}{+0.40} & \textcolor{gaingreen}{+1.48} & \textcolor{gaingreen}{+0.35} & \textcolor{gaingreen}{+1.97} & \textcolor{gaingreen}{+2.79} & \textcolor{gaingreen}{+3.99} & \textcolor{gaingreen}{+2.33} \\
\midrule
\multirow{3}{*}{LUAD (CPTAC, ext.)}
 & Base  & 73.59$_{(1.44)}$ & 74.19$_{(2.07)}$ & 65.54$_{(5.18)}$ & 53.53$_{(5.17)}$ & 53.73$_{(5.31)}$ & 73.76$_{(1.98)}$ & 49.76$_{(1.43)}$ & 74.31$_{(4.05)}$ & 64.80$_{(10.08)}$ \\
 & +Ours & 73.69$_{(3.01)}$ & 74.68$_{(2.58)}$ & 71.70$_{(1.89)}$ & 66.20$_{(3.22)}$ & 67.61$_{(4.29)}$ & 74.21$_{(2.51)}$ & 67.07$_{(5.40)}$ & 74.70$_{(3.89)}$ & 71.23$_{(3.44)}$ \\
 & $\Delta$ & \textcolor{gaingreen}{+0.10} & \textcolor{gaingreen}{+0.49} & \textcolor{gaingreen}{+6.16} & \textcolor{gaingreen}{+12.67} & \textcolor{gaingreen}{+13.88} & \textcolor{gaingreen}{+0.45} & \textcolor{gaingreen}{+17.31} & \textcolor{gaingreen}{+0.39} & \textcolor{gaingreen}{+6.43} \\
\midrule
\multirow{3}{*}{LUAD (NLST, ext.)}
 & Base  & 63.53$_{(0.61)}$ & 62.05$_{(1.27)}$ & 53.70$_{(1.24)}$ & 56.12$_{(0.94)}$ & 51.84$_{(2.09)}$ & 64.69$_{(1.23)}$ & 46.17$_{(3.03)}$ & 61.37$_{(1.40)}$ & 57.43$_{(6.13)}$ \\
 & +Ours & 63.84$_{(1.77)}$ & 62.52$_{(4.05)}$ & 61.35$_{(2.66)}$ & 60.49$_{(1.64)}$ & 52.33$_{(3.60)}$ & 65.32$_{(1.52)}$ & 54.11$_{(2.31)}$ & 61.65$_{(2.73)}$ & 60.20$_{(4.29)}$ \\
 & $\Delta$ & \textcolor{gaingreen}{+0.31} & \textcolor{gaingreen}{+0.47} & \textcolor{gaingreen}{+7.66} & \textcolor{gaingreen}{+4.38} & \textcolor{gaingreen}{+0.49} & \textcolor{gaingreen}{+0.63} & \textcolor{gaingreen}{+7.95} & \textcolor{gaingreen}{+0.28} & \textcolor{gaingreen}{+2.77} \\
\midrule
\multirow{3}{*}{LUSC (TCGA)}
 & Base  & 51.63$_{(4.05)}$ & 49.88$_{(1.87)}$ & 54.15$_{(7.06)}$ & 53.57$_{(2.32)}$ & 43.00$_{(5.55)}$ & 50.42$_{(4.45)}$ & 45.45$_{(6.75)}$ & 47.50$_{(2.90)}$ & 49.45$_{(3.65)}$ \\
 & +Ours & 51.82$_{(3.72)}$ & 50.39$_{(3.20)}$ & 57.48$_{(6.80)}$ & 56.28$_{(9.06)}$ & 51.99$_{(5.57)}$ & 51.46$_{(4.62)}$ & 52.06$_{(1.59)}$ & 54.69$_{(7.37)}$ & 53.27$_{(2.39)}$ \\
 & $\Delta$ & \textcolor{gaingreen}{+0.20} & \textcolor{gaingreen}{+0.51} & \textcolor{gaingreen}{+3.33} & \textcolor{gaingreen}{+2.71} & \textcolor{gaingreen}{+8.99} & \textcolor{gaingreen}{+1.04} & \textcolor{gaingreen}{+6.61} & \textcolor{gaingreen}{+7.19} & \textcolor{gaingreen}{+3.82} \\
\midrule
\multirow{3}{*}{LUSC (CPTAC, ext.)}
 & Base  & 65.26$_{(2.27)}$ & 65.42$_{(1.97)}$ & 48.02$_{(4.04)}$ & 50.96$_{(3.15)}$ & 51.30$_{(4.90)}$ & 67.96$_{(0.76)}$ & 46.04$_{(5.32)}$ & 62.48$_{(3.27)}$ & 57.18$_{(8.36)}$ \\
 & +Ours & 68.45$_{(2.60)}$ & 70.78$_{(0.51)}$ & 67.24$_{(2.02)}$ & 65.94$_{(1.37)}$ & 53.56$_{(3.67)}$ & 68.05$_{(2.46)}$ & 60.37$_{(6.12)}$ & 67.71$_{(4.47)}$ & 65.26$_{(5.24)}$ \\
 & $\Delta$ & \textcolor{gaingreen}{+3.19} & \textcolor{gaingreen}{+5.36} & \textcolor{gaingreen}{+19.23} & \textcolor{gaingreen}{+14.98} & \textcolor{gaingreen}{+2.26} & \textcolor{gaingreen}{+0.09} & \textcolor{gaingreen}{+14.33} & \textcolor{gaingreen}{+5.23} & \textcolor{gaingreen}{+8.09} \\
\midrule
\multirow{3}{*}{LUSC (NLST, ext.)}
 & Base  & 69.08$_{(1.05)}$ & 68.50$_{(1.58)}$ & 59.19$_{(2.20)}$ & 59.61$_{(3.43)}$ & 53.57$_{(8.89)}$ & 67.01$_{(0.89)}$ & 49.94$_{(0.64)}$ & 63.64$_{(0.92)}$ & 61.32$_{(6.58)}$ \\
 & +Ours & 69.96$_{(2.09)}$ & 69.27$_{(1.69)}$ & 62.90$_{(1.89)}$ & 62.37$_{(2.61)}$ & 70.60$_{(2.07)}$ & 68.14$_{(2.13)}$ & 50.48$_{(3.41)}$ & 62.05$_{(1.43)}$ & 64.47$_{(6.25)}$ \\
 & $\Delta$ & \textcolor{gaingreen}{+0.88} & \textcolor{gaingreen}{+0.77} & \textcolor{gaingreen}{+3.70} & \textcolor{gaingreen}{+2.76} & \textcolor{gaingreen}{+17.03} & \textcolor{gaingreen}{+1.13} & \textcolor{gaingreen}{+0.53} & \textcolor{lossred}{-1.59} & \textcolor{gaingreen}{+3.15} \\
\midrule
\multirow{3}{*}{COAD (TCGA)}
 & Base  & 60.71$_{(8.60)}$ & 59.84$_{(9.68)}$ & 60.20$_{(11.62)}$ & 58.87$_{(5.45)}$ & 48.38$_{(3.52)}$ & 58.70$_{(10.26)}$ & 48.92$_{(11.55)}$ & 57.22$_{(11.16)}$ & 56.61$_{(4.70)}$ \\
 & +Ours & 62.08$_{(7.98)}$ & 63.02$_{(6.78)}$ & 63.73$_{(8.92)}$ & 65.00$_{(2.67)}$ & 61.57$_{(2.79)}$ & 60.10$_{(9.13)}$ & 64.96$_{(2.16)}$ & 69.34$_{(2.21)}$ & 63.73$_{(2.64)}$ \\
 & $\Delta$ & \textcolor{gaingreen}{+1.38} & \textcolor{gaingreen}{+3.18} & \textcolor{gaingreen}{+3.53} & \textcolor{gaingreen}{+6.13} & \textcolor{gaingreen}{+13.20} & \textcolor{gaingreen}{+1.40} & \textcolor{gaingreen}{+16.04} & \textcolor{gaingreen}{+12.12} & \textcolor{gaingreen}{+7.12} \\
\midrule
\multirow{3}{*}{COAD (CPTAC, ext.)}
 & Base  & 60.72$_{(1.54)}$ & 61.69$_{(0.61)}$ & 59.04$_{(2.91)}$ & 53.12$_{(2.54)}$ & 48.31$_{(3.08)}$ & 62.18$_{(0.92)}$ & 51.01$_{(1.77)}$ & 58.01$_{(3.31)}$ & 56.76$_{(4.92)}$ \\
 & +Ours & 62.88$_{(1.16)}$ & 62.33$_{(1.40)}$ & 65.33$_{(1.22)}$ & 55.97$_{(2.50)}$ & 55.93$_{(2.55)}$ & 62.88$_{(0.67)}$ & 51.60$_{(1.57)}$ & 61.11$_{(2.32)}$ & 59.75$_{(4.40)}$ \\
 & $\Delta$ & \textcolor{gaingreen}{+2.16} & \textcolor{gaingreen}{+0.64} & \textcolor{gaingreen}{+6.29} & \textcolor{gaingreen}{+2.85} & \textcolor{gaingreen}{+7.63} & \textcolor{gaingreen}{+0.70} & \textcolor{gaingreen}{+0.58} & \textcolor{gaingreen}{+3.10} & \textcolor{gaingreen}{+3.00} \\
\bottomrule
\end{tabular}}
\end{table}

\begin{table}[t]
\centering
\caption{\textbf{Molecular biomarker prediction (detailed breakdown).} Results across 15 biomarker prediction tasks spanning GBMLGG, BCNB, BRCA, and Lung cohorts using eight MIL aggregators. TCGA and EBRAINS are reported separately for GBMLGG. AUROC is used for binary classification tasks and balanced accuracy for Histomol. Matched 5-fold CV splits for Base and MIST (10-fold for BCNB); standard errors across folds are in parentheses. Avg. denotes the cross-aggregator mean performance.}
\label{tab:biomarker_detailed}
\resizebox{\textwidth}{!}{%
\begin{tabular}{llccccccccc}
\toprule
Task & Status & ABMIL & CLAM & TransMIL & Trans. & ILRA & Mean & Max & DSMIL & Avg. \\
\midrule
\multirow{3}{*}{GBMLGG IDH1 (TCGA)}
 & Base  & 95.86$_{(0.38)}$ & 95.94$_{(0.99)}$ & 94.10$_{(1.37)}$ & 95.34$_{(1.56)}$ & 94.11$_{(2.20)}$ & 96.24$_{(0.90)}$ & 92.42$_{(3.56)}$ & 95.12$_{(0.71)}$ & 94.89$_{(1.20)}$ \\
 & +Ours & 96.61$_{(0.82)}$ & 96.61$_{(0.59)}$ & 95.63$_{(1.43)}$ & 96.38$_{(0.81)}$ & 96.66$_{(1.36)}$ & 96.57$_{(0.80)}$ & 94.63$_{(1.29)}$ & 96.62$_{(0.54)}$ & 96.21$_{(0.68)}$ \\
 & $\Delta$ & \textcolor{gaingreen}{+0.76} & \textcolor{gaingreen}{+0.68} & \textcolor{gaingreen}{+1.53} & \textcolor{gaingreen}{+1.03} & \textcolor{gaingreen}{+2.55} & \textcolor{gaingreen}{+0.33} & \textcolor{gaingreen}{+2.21} & \textcolor{gaingreen}{+1.50} & \textcolor{gaingreen}{+1.32} \\
\midrule
\multirow{3}{*}{GBMLGG IDH1 (EBRAINS)}
 & Base  & 93.40$_{(2.23)}$ & 94.23$_{(1.03)}$ & 89.91$_{(2.15)}$ & 93.09$_{(1.58)}$ & 88.66$_{(1.95)}$ & 91.68$_{(1.21)}$ & 90.33$_{(1.12)}$ & 92.39$_{(2.85)}$ & 91.71$_{(1.80)}$ \\
 & +Ours & 94.95$_{(0.81)}$ & 94.87$_{(1.17)}$ & 93.27$_{(1.81)}$ & 94.81$_{(1.10)}$ & 91.68$_{(1.86)}$ & 93.29$_{(0.99)}$ & 92.36$_{(0.85)}$ & 94.82$_{(0.97)}$ & 93.76$_{(1.21)}$ \\
 & $\Delta$ & \textcolor{gaingreen}{+1.55} & \textcolor{gaingreen}{+0.64} & \textcolor{gaingreen}{+3.36} & \textcolor{gaingreen}{+1.73} & \textcolor{gaingreen}{+3.02} & \textcolor{gaingreen}{+1.62} & \textcolor{gaingreen}{+2.03} & \textcolor{gaingreen}{+2.43} & \textcolor{gaingreen}{+2.04} \\
\midrule
\multirow{3}{*}{GBMLGG Histomol (TCGA)}
 & Base  & 66.57$_{(6.03)}$ & 66.37$_{(5.01)}$ & 63.17$_{(5.89)}$ & 66.74$_{(3.45)}$ & 56.18$_{(5.86)}$ & 66.57$_{(4.91)}$ & 60.10$_{(5.61)}$ & 64.92$_{(8.24)}$ & 63.83$_{(3.61)}$ \\
 & +Ours & 68.85$_{(5.92)}$ & 69.01$_{(5.15)}$ & 66.65$_{(6.12)}$ & 68.32$_{(3.56)}$ & 58.60$_{(5.91)}$ & 67.86$_{(4.63)}$ & 61.65$_{(5.01)}$ & 65.94$_{(5.39)}$ & 65.86$_{(3.54)}$ \\
 & $\Delta$ & \textcolor{gaingreen}{+2.28} & \textcolor{gaingreen}{+2.64} & \textcolor{gaingreen}{+3.48} & \textcolor{gaingreen}{+1.58} & \textcolor{gaingreen}{+2.41} & \textcolor{gaingreen}{+1.29} & \textcolor{gaingreen}{+1.54} & \textcolor{gaingreen}{+1.02} & \textcolor{gaingreen}{+2.03} \\
\midrule
\multirow{3}{*}{GBMLGG Histomol (EBRAINS)}
 & Base  & 54.69$_{(2.40)}$ & 54.68$_{(2.72)}$ & 50.24$_{(5.07)}$ & 49.05$_{(3.75)}$ & 42.19$_{(2.25)}$ & 49.63$_{(2.86)}$ & 49.79$_{(1.23)}$ & 52.57$_{(3.41)}$ & 50.35$_{(3.74)}$ \\
 & +Ours & 56.71$_{(3.39)}$ & 56.70$_{(2.84)}$ & 53.77$_{(3.26)}$ & 52.97$_{(4.63)}$ & 46.09$_{(1.39)}$ & 51.48$_{(3.47)}$ & 53.32$_{(1.30)}$ & 54.25$_{(2.59)}$ & 53.16$_{(3.15)}$ \\
 & $\Delta$ & \textcolor{gaingreen}{+2.02} & \textcolor{gaingreen}{+2.02} & \textcolor{gaingreen}{+3.54} & \textcolor{gaingreen}{+3.92} & \textcolor{gaingreen}{+3.90} & \textcolor{gaingreen}{+1.85} & \textcolor{gaingreen}{+3.54} & \textcolor{gaingreen}{+1.68} & \textcolor{gaingreen}{+2.81} \\
\midrule
\multirow{3}{*}{BCNB ER}
 & Base  & 92.38$_{(2.18)}$ & 93.07$_{(1.89)}$ & 90.66$_{(3.18)}$ & 90.98$_{(3.87)}$ & 87.29$_{(4.57)}$ & 88.93$_{(2.85)}$ & 75.67$_{(8.21)}$ & 91.84$_{(2.92)}$ & 88.85$_{(5.28)}$ \\
 & +Ours & 93.31$_{(2.34)}$ & 92.80$_{(2.02)}$ & 92.37$_{(2.90)}$ & 93.13$_{(1.73)}$ & 90.23$_{(3.02)}$ & 90.50$_{(2.01)}$ & 91.29$_{(3.17)}$ & 92.81$_{(2.81)}$ & 92.06$_{(1.13)}$ \\
 & $\Delta$ & \textcolor{gaingreen}{+0.92} & \textcolor{lossred}{$-$0.27} & \textcolor{gaingreen}{+1.71} & \textcolor{gaingreen}{+2.16} & \textcolor{gaingreen}{+2.95} & \textcolor{gaingreen}{+1.57} & \textcolor{gaingreen}{+15.61} & \textcolor{gaingreen}{+0.98} & \textcolor{gaingreen}{+3.20} \\
\midrule
\multirow{3}{*}{BCNB HER2}
 & Base  & 76.54$_{(4.93)}$ & 76.92$_{(4.44)}$ & 73.56$_{(5.31)}$ & 74.34$_{(4.89)}$ & 70.21$_{(6.42)}$ & 72.70$_{(5.03)}$ & 64.10$_{(6.82)}$ & 74.73$_{(3.73)}$ & 72.89$_{(3.87)}$ \\
 & +Ours & 77.87$_{(4.20)}$ & 77.80$_{(4.97)}$ & 72.81$_{(5.69)}$ & 77.51$_{(3.83)}$ & 72.58$_{(6.12)}$ & 74.26$_{(4.15)}$ & 73.57$_{(3.94)}$ & 75.53$_{(4.39)}$ & 75.24$_{(2.10)}$ \\
 & $\Delta$ & \textcolor{gaingreen}{+1.33} & \textcolor{gaingreen}{+0.88} & \textcolor{lossred}{$-$0.75} & \textcolor{gaingreen}{+3.17} & \textcolor{gaingreen}{+2.37} & \textcolor{gaingreen}{+1.56} & \textcolor{gaingreen}{+9.47} & \textcolor{gaingreen}{+0.80} & \textcolor{gaingreen}{+2.35} \\
\midrule
\multirow{3}{*}{BCNB PR}
 & Base  & 86.42$_{(3.12)}$ & 87.06$_{(2.91)}$ & 83.74$_{(4.80)}$ & 84.75$_{(3.56)}$ & 79.48$_{(5.53)}$ & 83.87$_{(4.32)}$ & 80.77$_{(7.23)}$ & 86.44$_{(3.21)}$ & 84.07$_{(2.57)}$ \\
 & +Ours & 87.97$_{(2.96)}$ & 88.09$_{(2.40)}$ & 86.32$_{(3.53)}$ & 87.43$_{(3.75)}$ & 83.29$_{(3.98)}$ & 85.18$_{(4.12)}$ & 84.77$_{(3.55)}$ & 86.18$_{(3.26)}$ & 86.15$_{(1.57)}$ \\
 & $\Delta$ & \textcolor{gaingreen}{+1.54} & \textcolor{gaingreen}{+1.03} & \textcolor{gaingreen}{+2.59} & \textcolor{gaingreen}{+2.67} & \textcolor{gaingreen}{+3.81} & \textcolor{gaingreen}{+1.31} & \textcolor{gaingreen}{+4.00} & \textcolor{lossred}{$-$0.26} & \textcolor{gaingreen}{+2.09} \\
\midrule
\multirow{3}{*}{BRCA ER}
 & Base  & 90.61$_{(2.33)}$ & 91.29$_{(2.93)}$ & 86.30$_{(5.68)}$ & 86.42$_{(2.83)}$ & 80.78$_{(3.84)}$ & 87.95$_{(2.49)}$ & 86.13$_{(3.46)}$ & 89.21$_{(3.60)}$ & 87.34$_{(3.09)}$ \\
 & +Ours & 91.19$_{(2.50)}$ & 90.56$_{(3.38)}$ & 89.45$_{(2.68)}$ & 90.84$_{(2.84)}$ & 85.45$_{(2.09)}$ & 88.66$_{(2.26)}$ & 88.40$_{(2.14)}$ & 89.87$_{(2.21)}$ & 89.30$_{(1.73)}$ \\
 & $\Delta$ & \textcolor{gaingreen}{+0.58} & \textcolor{lossred}{$-$0.73} & \textcolor{gaingreen}{+3.15} & \textcolor{gaingreen}{+4.42} & \textcolor{gaingreen}{+4.67} & \textcolor{gaingreen}{+0.70} & \textcolor{gaingreen}{+2.27} & \textcolor{gaingreen}{+0.66} & \textcolor{gaingreen}{+1.97} \\
\midrule
\multirow{3}{*}{BRCA HER2}
 & Base  & 70.43$_{(7.29)}$ & 70.06$_{(5.50)}$ & 64.92$_{(4.48)}$ & 63.77$_{(4.87)}$ & 57.67$_{(3.70)}$ & 66.20$_{(6.81)}$ & 60.12$_{(4.89)}$ & 62.13$_{(5.79)}$ & 64.41$_{(4.20)}$ \\
 & +Ours & 72.20$_{(4.06)}$ & 68.67$_{(6.03)}$ & 69.35$_{(4.21)}$ & 71.00$_{(2.23)}$ & 65.40$_{(6.18)}$ & 68.27$_{(5.69)}$ & 67.45$_{(4.97)}$ & 68.59$_{(6.87)}$ & 68.87$_{(1.95)}$ \\
 & $\Delta$ & \textcolor{gaingreen}{+1.77} & \textcolor{lossred}{$-$1.40} & \textcolor{gaingreen}{+4.44} & \textcolor{gaingreen}{+7.24} & \textcolor{gaingreen}{+7.73} & \textcolor{gaingreen}{+2.07} & \textcolor{gaingreen}{+7.34} & \textcolor{gaingreen}{+6.46} & \textcolor{gaingreen}{+4.45} \\
\midrule
\multirow{3}{*}{BRCA PIK3CA}
 & Base  & 64.43$_{(3.80)}$ & 65.20$_{(5.88)}$ & 57.88$_{(4.37)}$ & 59.11$_{(5.37)}$ & 57.36$_{(4.74)}$ & 61.70$_{(4.82)}$ & 58.73$_{(6.49)}$ & 61.71$_{(6.16)}$ & 60.77$_{(2.78)}$ \\
 & +Ours & 64.37$_{(5.11)}$ & 65.94$_{(4.38)}$ & 63.68$_{(3.89)}$ & 63.19$_{(4.60)}$ & 60.88$_{(5.00)}$ & 62.71$_{(3.69)}$ & 62.06$_{(3.88)}$ & 62.63$_{(4.96)}$ & 63.18$_{(1.43)}$ \\
 & $\Delta$ & \textcolor{lossred}{$-$0.07} & \textcolor{gaingreen}{+0.74} & \textcolor{gaingreen}{+5.80} & \textcolor{gaingreen}{+4.08} & \textcolor{gaingreen}{+3.52} & \textcolor{gaingreen}{+1.02} & \textcolor{gaingreen}{+3.34} & \textcolor{gaingreen}{+0.92} & \textcolor{gaingreen}{+2.42} \\
\midrule
\multirow{3}{*}{BRCA PR}
 & Base  & 79.91$_{(4.18)}$ & 80.24$_{(2.81)}$ & 74.70$_{(5.53)}$ & 74.06$_{(2.54)}$ & 69.77$_{(3.75)}$ & 77.48$_{(3.36)}$ & 72.67$_{(2.98)}$ & 78.41$_{(3.96)}$ & 75.91$_{(3.47)}$ \\
 & +Ours & 80.89$_{(2.99)}$ & 80.36$_{(3.01)}$ & 78.52$_{(4.46)}$ & 77.54$_{(4.82)}$ & 73.96$_{(2.58)}$ & 77.64$_{(3.86)}$ & 78.54$_{(4.62)}$ & 78.20$_{(2.03)}$ & 78.21$_{(1.96)}$ \\
 & $\Delta$ & \textcolor{gaingreen}{+0.97} & \textcolor{gaingreen}{+0.12} & \textcolor{gaingreen}{+3.82} & \textcolor{gaingreen}{+3.48} & \textcolor{gaingreen}{+4.19} & \textcolor{gaingreen}{+0.16} & \textcolor{gaingreen}{+5.87} & \textcolor{lossred}{$-$0.21} & \textcolor{gaingreen}{+2.30} \\
\midrule
\multirow{3}{*}{Lung EGFR}
 & Base  & 71.71$_{(7.74)}$ & 69.68$_{(7.70)}$ & 70.08$_{(5.08)}$ & 68.24$_{(7.19)}$ & 67.40$_{(7.64)}$ & 72.02$_{(4.50)}$ & 59.65$_{(7.54)}$ & 65.72$_{(8.62)}$ & 68.06$_{(3.75)}$ \\
 & +Ours & 73.86$_{(3.72)}$ & 73.37$_{(3.22)}$ & 69.39$_{(8.39)}$ & 72.37$_{(5.63)}$ & 74.71$_{(3.01)}$ & 72.68$_{(3.87)}$ & 65.18$_{(5.40)}$ & 71.21$_{(5.74)}$ & 71.60$_{(2.87)}$ \\
 & $\Delta$ & \textcolor{gaingreen}{+2.15} & \textcolor{gaingreen}{+3.68} & \textcolor{lossred}{$-$0.69} & \textcolor{gaingreen}{+4.13} & \textcolor{gaingreen}{+7.32} & \textcolor{gaingreen}{+0.67} & \textcolor{gaingreen}{+5.53} & \textcolor{gaingreen}{+5.49} & \textcolor{gaingreen}{+3.53} \\
\midrule
\multirow{3}{*}{Lung KRAS}
 & Base  & 82.44$_{(4.21)}$ & 82.71$_{(3.57)}$ & 79.13$_{(4.72)}$ & 79.53$_{(1.96)}$ & 72.22$_{(2.62)}$ & 82.38$_{(3.26)}$ & 78.76$_{(5.59)}$ & 82.79$_{(3.37)}$ & 80.00$_{(3.35)}$ \\
 & +Ours & 83.34$_{(3.13)}$ & 83.60$_{(3.53)}$ & 82.86$_{(2.66)}$ & 82.52$_{(4.24)}$ & 79.05$_{(2.45)}$ & 82.21$_{(3.94)}$ & 81.53$_{(1.62)}$ & 81.28$_{(5.95)}$ & 82.05$_{(1.36)}$ \\
 & $\Delta$ & \textcolor{gaingreen}{+0.90} & \textcolor{gaingreen}{+0.88} & \textcolor{gaingreen}{+3.73} & \textcolor{gaingreen}{+3.00} & \textcolor{gaingreen}{+6.84} & \textcolor{lossred}{$-$0.17} & \textcolor{gaingreen}{+2.78} & \textcolor{lossred}{$-$1.51} & \textcolor{gaingreen}{+2.05} \\
\midrule
\multirow{3}{*}{Lung STK11}
 & Base  & 85.82$_{(2.52)}$ & 85.35$_{(2.74)}$ & 80.42$_{(2.26)}$ & 77.26$_{(9.01)}$ & 70.48$_{(3.59)}$ & 80.62$_{(2.44)}$ & 61.72$_{(9.41)}$ & 81.89$_{(3.70)}$ & 77.94$_{(7.63)}$ \\
 & +Ours & 86.62$_{(3.87)}$ & 87.17$_{(2.98)}$ & 84.95$_{(5.08)}$ & 83.39$_{(5.90)}$ & 78.46$_{(5.16)}$ & 81.67$_{(4.29)}$ & 78.40$_{(6.13)}$ & 83.48$_{(3.79)}$ & 83.02$_{(3.13)}$ \\
 & $\Delta$ & \textcolor{gaingreen}{+0.80} & \textcolor{gaingreen}{+1.82} & \textcolor{gaingreen}{+4.53} & \textcolor{gaingreen}{+6.13} & \textcolor{gaingreen}{+7.98} & \textcolor{gaingreen}{+1.05} & \textcolor{gaingreen}{+16.69} & \textcolor{gaingreen}{+1.58} & \textcolor{gaingreen}{+5.07} \\
\midrule
\multirow{3}{*}{Lung TP53}
 & Base  & 71.17$_{(3.79)}$ & 72.92$_{(2.62)}$ & 67.32$_{(4.09)}$ & 69.37$_{(4.65)}$ & 63.75$_{(4.36)}$ & 70.08$_{(3.58)}$ & 61.88$_{(5.61)}$ & 68.54$_{(4.64)}$ & 68.13$_{(3.47)}$ \\
 & +Ours & 73.36$_{(3.86)}$ & 73.62$_{(3.09)}$ & 71.83$_{(3.72)}$ & 73.59$_{(4.36)}$ & 68.80$_{(3.42)}$ & 72.08$_{(3.42)}$ & 66.89$_{(2.55)}$ & 71.49$_{(3.41)}$ & 71.46$_{(2.27)}$ \\
 & $\Delta$ & \textcolor{gaingreen}{+2.19} & \textcolor{gaingreen}{+0.70} & \textcolor{gaingreen}{+4.51} & \textcolor{gaingreen}{+4.22} & \textcolor{gaingreen}{+5.06} & \textcolor{gaingreen}{+2.00} & \textcolor{gaingreen}{+5.01} & \textcolor{gaingreen}{+2.95} & \textcolor{gaingreen}{+3.33} \\
\bottomrule
\end{tabular}}
\end{table}

\begin{table}[t]
\centering
\caption{\textbf{Tissue subtyping (detailed breakdown).} Results across NSCLC, PANDA, BRACS, and EBRAINS tissue subtyping tasks using eight MIL aggregators. NSCLC is evaluated separately on TCGA, CPTAC, and NLST cohorts. AUROC is reported for NSCLC, weighted $\kappa$ for PANDA, and balanced accuracy for BRACS and EBRAINS. NSCLC uses 5-fold CV; PANDA, BRACS, and EBRAINS use predefined splits with standard errors from 1{,}000 bootstrap resamples. Avg. denotes the cross-aggregator mean performance.}
\label{tab:classification_detailed}
\resizebox{\textwidth}{!}{%
\begin{tabular}{llccccccccc}
\toprule
Task & Status & ABMIL & CLAM & TransMIL & Trans. & ILRA & Mean & Max & DSMIL & Avg. \\
\midrule
\multirow{3}{*}{NSCLC (TCGA, AUROC)}
 & Base  & 98.25$_{(1.05)}$ & 98.31$_{(1.09)}$ & 98.04$_{(1.20)}$ & 97.89$_{(1.14)}$ & 96.96$_{(1.75)}$ & 96.95$_{(1.15)}$ & 96.50$_{(0.73)}$ & 97.70$_{(1.21)}$ & 97.57$_{(0.64)}$ \\
 & +Ours & 98.51$_{(0.98)}$ & 98.53$_{(0.94)}$ & 98.26$_{(1.04)}$ & 98.31$_{(1.01)}$ & 97.61$_{(1.35)}$ & 97.53$_{(1.14)}$ & 97.84$_{(0.77)}$ & 98.20$_{(0.94)}$ & 98.10$_{(0.36)}$ \\
 & $\Delta$ & \textcolor{gaingreen}{+0.26} & \textcolor{gaingreen}{+0.21} & \textcolor{gaingreen}{+0.22} & \textcolor{gaingreen}{+0.42} & \textcolor{gaingreen}{+0.65} & \textcolor{gaingreen}{+0.58} & \textcolor{gaingreen}{+1.35} & \textcolor{gaingreen}{+0.51} & \textcolor{gaingreen}{+0.52} \\
\midrule
\multirow{3}{*}{NSCLC (CPTAC, ext.)}
 & Base  & 97.20$_{(0.42)}$ & 97.50$_{(0.31)}$ & 93.42$_{(1.84)}$ & 94.19$_{(1.94)}$ & 92.09$_{(1.42)}$ & 93.95$_{(0.75)}$ & 96.13$_{(0.65)}$ & 96.71$_{(0.62)}$ & 95.15$_{(1.87)}$ \\
 & +Ours & 96.86$_{(0.53)}$ & 96.69$_{(0.46)}$ & 95.33$_{(1.15)}$ & 95.54$_{(1.45)}$ & 94.88$_{(1.18)}$ & 94.96$_{(0.80)}$ & 95.96$_{(0.56)}$ & 96.11$_{(0.62)}$ & 95.79$_{(0.70)}$ \\
 & $\Delta$ & \textcolor{lossred}{-0.34} & \textcolor{lossred}{-0.81} & \textcolor{gaingreen}{+1.92} & \textcolor{gaingreen}{+1.35} & \textcolor{gaingreen}{+2.78} & \textcolor{gaingreen}{+1.02} & \textcolor{lossred}{-0.16} & \textcolor{lossred}{-0.60} & \textcolor{gaingreen}{+0.64} \\
\midrule
\multirow{3}{*}{NSCLC (NLST, ext.)}
 & Base  & 86.32$_{(0.65)}$ & 87.56$_{(0.60)}$ & 85.12$_{(0.62)}$ & 86.90$_{(0.80)}$ & 78.21$_{(1.09)}$ & 80.10$_{(0.39)}$ & 88.47$_{(0.99)}$ & 85.87$_{(0.86)}$ & 84.82$_{(3.44)}$ \\
 & +Ours & 88.29$_{(0.58)}$ & 88.45$_{(0.63)}$ & 86.60$_{(0.94)}$ & 87.93$_{(0.65)}$ & 81.34$_{(1.43)}$ & 81.96$_{(0.83)}$ & 89.09$_{(0.79)}$ & 88.92$_{(0.63)}$ & 86.57$_{(2.93)}$ \\
 & $\Delta$ & \textcolor{gaingreen}{+1.97} & \textcolor{gaingreen}{+0.89} & \textcolor{gaingreen}{+1.48} & \textcolor{gaingreen}{+1.03} & \textcolor{gaingreen}{+3.13} & \textcolor{gaingreen}{+1.86} & \textcolor{gaingreen}{+0.61} & \textcolor{gaingreen}{+3.04} & \textcolor{gaingreen}{+1.75} \\
\midrule
\multirow{3}{*}{PANDA (Weighted $\kappa$)}
 & Base  & 93.40$_{(0.65)}$ & 93.72$_{(0.60)}$ & 93.42$_{(0.62)}$ & 92.70$_{(0.77)}$ & 92.94$_{(0.76)}$ & 87.74$_{(1.18)}$ & 83.05$_{(1.64)}$ & 94.24$_{(0.60)}$ & 91.40$_{(3.69)}$ \\
 & +Ours & 94.15$_{(0.69)}$ & 94.15$_{(0.59)}$ & 93.92$_{(0.63)}$ & 94.19$_{(0.60)}$ & 95.47$_{(0.48)}$ & 92.17$_{(0.85)}$ & 86.29$_{(1.32)}$ & 93.99$_{(0.73)}$ & 93.04$_{(2.68)}$ \\
 & $\Delta$ & \textcolor{gaingreen}{+0.75} & \textcolor{gaingreen}{+0.43} & \textcolor{gaingreen}{+0.50} & \textcolor{gaingreen}{+1.48} & \textcolor{gaingreen}{+2.53} & \textcolor{gaingreen}{+4.43} & \textcolor{gaingreen}{+3.25} & \textcolor{lossred}{-0.24} & \textcolor{gaingreen}{+1.64} \\
\midrule
\multirow{3}{*}{BRACS-C (Bal.\ acc.)}
 & Base  & 62.59$_{(5.18)}$ & 62.41$_{(5.13)}$ & 65.99$_{(4.21)}$ & 61.82$_{(4.71)}$ & 44.70$_{(5.34)}$ & 54.89$_{(5.38)}$ & 60.28$_{(5.02)}$ & 57.61$_{(5.19)}$ & 58.79$_{(6.18)}$ \\
 & +Ours & 72.78$_{(4.79)}$ & 70.29$_{(4.90)}$ & 70.70$_{(4.97)}$ & 72.60$_{(4.67)}$ & 51.40$_{(5.14)}$ & 58.42$_{(5.49)}$ & 65.31$_{(5.48)}$ & 59.87$_{(5.14)}$ & 65.17$_{(7.36)}$ \\
 & $\Delta$ & \textcolor{gaingreen}{+10.19} & \textcolor{gaingreen}{+7.88} & \textcolor{gaingreen}{+4.71} & \textcolor{gaingreen}{+10.78} & \textcolor{gaingreen}{+6.70} & \textcolor{gaingreen}{+3.53} & \textcolor{gaingreen}{+5.03} & \textcolor{gaingreen}{+2.26} & \textcolor{gaingreen}{+6.39} \\
\midrule
\multirow{3}{*}{BRACS-F (Bal.\ acc.)}
 & Base  & 36.66$_{(4.70)}$ & 44.02$_{(4.60)}$ & 38.98$_{(4.79)}$ & 41.15$_{(5.17)}$ & 29.93$_{(4.60)}$ & 31.14$_{(4.60)}$ & 28.41$_{(4.24)}$ & 34.59$_{(5.19)}$ & 35.61$_{(5.23)}$ \\
 & +Ours & 44.35$_{(4.97)}$ & 45.22$_{(5.14)}$ & 45.58$_{(5.22)}$ & 41.90$_{(4.58)}$ & 30.25$_{(4.24)}$ & 32.20$_{(4.82)}$ & 36.02$_{(4.52)}$ & 36.36$_{(4.91)}$ & 38.98$_{(5.68)}$ \\
 & $\Delta$ & \textcolor{gaingreen}{+7.69} & \textcolor{gaingreen}{+1.20} & \textcolor{gaingreen}{+6.60} & \textcolor{gaingreen}{+0.75} & \textcolor{gaingreen}{+0.31} & \textcolor{gaingreen}{+1.06} & \textcolor{gaingreen}{+7.61} & \textcolor{gaingreen}{+1.77} & \textcolor{gaingreen}{+3.37} \\
\midrule
\multirow{3}{*}{EBRAINS-C (Bal.\ acc.)}
 & Base  & 90.77$_{(1.92)}$ & 89.86$_{(1.97)}$ & 87.97$_{(2.02)}$ & 88.11$_{(2.04)}$ & 84.74$_{(2.39)}$ & 86.97$_{(2.21)}$ & 75.91$_{(2.35)}$ & 83.24$_{(2.36)}$ & 85.95$_{(4.44)}$ \\
 & +Ours & 91.86$_{(1.84)}$ & 92.15$_{(1.79)}$ & 90.41$_{(1.94)}$ & 91.74$_{(1.68)}$ & 86.17$_{(2.25)}$ & 90.36$_{(2.02)}$ & 83.02$_{(2.26)}$ & 90.34$_{(2.04)}$ & 89.51$_{(3.02)}$ \\
 & $\Delta$ & \textcolor{gaingreen}{+1.08} & \textcolor{gaingreen}{+2.28} & \textcolor{gaingreen}{+2.45} & \textcolor{gaingreen}{+3.64} & \textcolor{gaingreen}{+1.43} & \textcolor{gaingreen}{+3.39} & \textcolor{gaingreen}{+7.12} & \textcolor{gaingreen}{+7.10} & \textcolor{gaingreen}{+3.56} \\
\midrule
\multirow{3}{*}{EBRAINS-F (Bal.\ acc.)}
 & Base  & 69.36$_{(2.20)}$ & 71.04$_{(2.19)}$ & 72.76$_{(2.00)}$ & 72.10$_{(1.96)}$ & 63.46$_{(2.21)}$ & 69.08$_{(2.20)}$ & 60.23$_{(2.28)}$ & 64.25$_{(2.20)}$ & 67.79$_{(4.28)}$ \\
 & +Ours & 74.87$_{(2.07)}$ & 74.15$_{(1.97)}$ & 73.82$_{(2.08)}$ & 75.38$_{(1.97)}$ & 68.12$_{(2.17)}$ & 73.79$_{(2.09)}$ & 63.78$_{(2.17)}$ & 68.36$_{(2.18)}$ & 71.53$_{(3.95)}$ \\
 & $\Delta$ & \textcolor{gaingreen}{+5.51} & \textcolor{gaingreen}{+3.11} & \textcolor{gaingreen}{+1.06} & \textcolor{gaingreen}{+3.28} & \textcolor{gaingreen}{+4.66} & \textcolor{gaingreen}{+4.70} & \textcolor{gaingreen}{+3.54} & \textcolor{gaingreen}{+4.11} & \textcolor{gaingreen}{+3.75} \\
\bottomrule
\end{tabular}}
\end{table}

\section{Implementation Details}
\label{app:implementation}

All whole-slide images are processed at $20\times$ magnification ($0.5\,\mu$m/pixel) and tiled into non-overlapping $256{\times}256$ patches. Patch embeddings are extracted using the frozen UNI2-h encoder~\citep{chen2024uni}, a ViT-H/14 pathology foundation model pretrained with the DINOv2 objective, producing $1{,}536$-dimensional feature vectors. Molecular prototypes are constructed once per organ from paired ST and H\&E data following Section~\ref{sec:prototype_construction}. Unless otherwise specified, MIST uses a projection dimension of $d{=}512$ and a low-rank bottleneck dimension of $r{=}16$.

All models are optimized using AdamW with learning rate $10^{-4}$ and weight decay $10^{-5}$. Training uses a batch size of $1$ with gradient accumulation over $32$ iterations, together with input feature dropout of $0.1$ and model dropout of $0.25$. A cosine learning rate schedule with one epoch of warmup is applied throughout training.

For classification tasks, models are trained for up to $20$ epochs with early stopping based on validation performance (patience $5$, minimum training length $10$ epochs). For survival prediction, all models are trained for $20$ epochs without early stopping. Unless otherwise noted, all MIL aggregators follow their published hyperparameter settings. All experiments are conducted on NVIDIA H100 GPUs.

\section{Baseline MIL Aggregators}
\label{app:baselines}

We evaluate MIST as a plug-in replacement for the standard projection layer across eight representative MIL aggregators, following the benchmark configuration of \citep{shao2026mammoth}. These aggregators span three broad categories: pooling-based methods, attention-based methods, and transformer-based architectures.

\textbf{MeanMIL} and \textbf{MaxMIL} are parameter-free pooling baselines that aggregate patch embeddings using mean pooling and max pooling, respectively. Despite their simplicity, they provide useful reference points for evaluating the contribution of learned aggregation mechanisms.

\textbf{ABMIL}~\citep{ilse2018attention} aggregates patch embeddings through gated attention, assigning each patch a learnable relevance weight before weighted pooling. Its simplicity and interpretability have made it one of the most widely used MIL baselines in computational pathology.

\textbf{CLAM}~\citep{lu2021data} extends ABMIL with a clustering-based auxiliary objective that encourages the attention mechanism to focus on diagnostically informative regions. This auxiliary supervision improves instance selection without requiring patch-level annotations.

\textbf{TransMIL}~\citep{shao2021transmil} introduces transformer-based aggregation with morphology-aware positional encoding to model long-range spatial dependencies among patches. By explicitly capturing inter-patch interactions, it provides richer contextual representations than independent attention pooling.

\textbf{TransformerMIL}~\citep{wagner2023transformer,vaswani2017attention} applies a standard transformer encoder without tissue-specific positional encoding, serving as a simplified variant of TransMIL that isolates the contribution of transformer-based sequence modeling.

\textbf{ILRA}~\citep{xiang2023exploring} formulates MIL aggregation through iterative low-rank approximation of the patch interaction matrix, constraining the representation to a compact latent subspace. This design reduces overfitting while preserving dominant patterns of inter-patch variation.

\textbf{DSMIL}~\citep{li2021dual} adopts a dual-stream architecture in which one branch identifies the most informative instance while the second branch propagates attention scores according to patch similarity to this anchor. The combination of instance-level and bag-level objectives improves robustness and calibration.

\section{More Interpretability Results}
\label{app:more_interp}

We provide additional interpretability visualizations on three held-out HEST slides from brain, breast, and lung tissues. For each slide, we visualize the whole-slide prototype affinity distributions predicted from H\&E alone, together with representative molecular prototypes and their associated spatial patterns.

\begin{figure*}[t]
    \centering
    \includegraphics[width=0.95\textwidth]{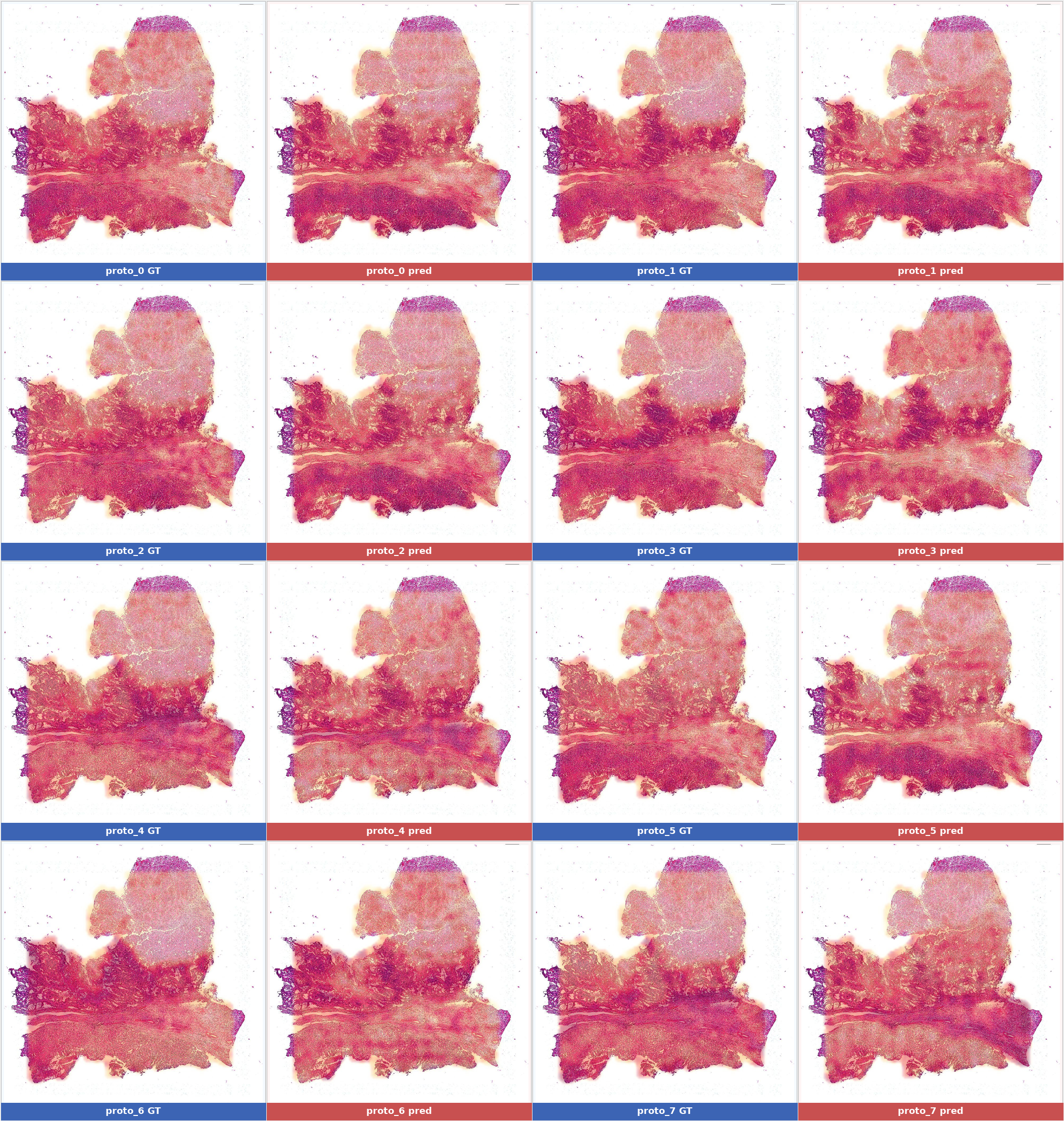}
    \caption{\textbf{Additional interpretability results on a held-out brain HEST slide.} Whole-slide prototype affinity visualizations generated from H\&E images without transcriptomics at inference.}
    \label{fig:interp_brain}
\end{figure*}

\begin{figure*}[t]
    \centering
    \includegraphics[width=0.95\textwidth]{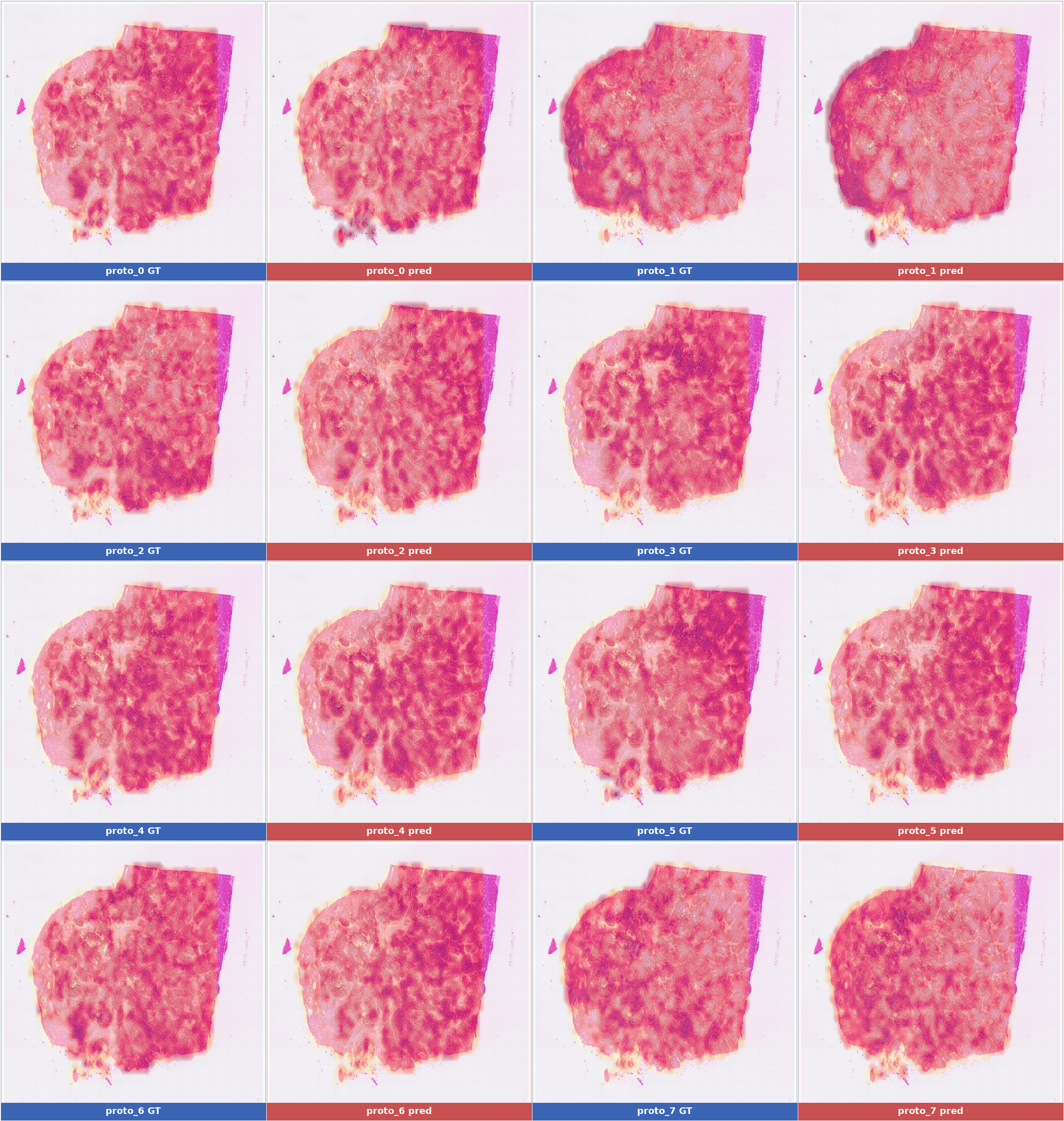}
    \caption{\textbf{Additional interpretability results on a held-out breast HEST slide.} Whole-slide prototype affinity visualizations generated from H\&E images without transcriptomics at inference.}
    \label{fig:interp_breast}
\end{figure*}

\begin{figure*}[t]
    \centering
    \includegraphics[width=0.95\textwidth]{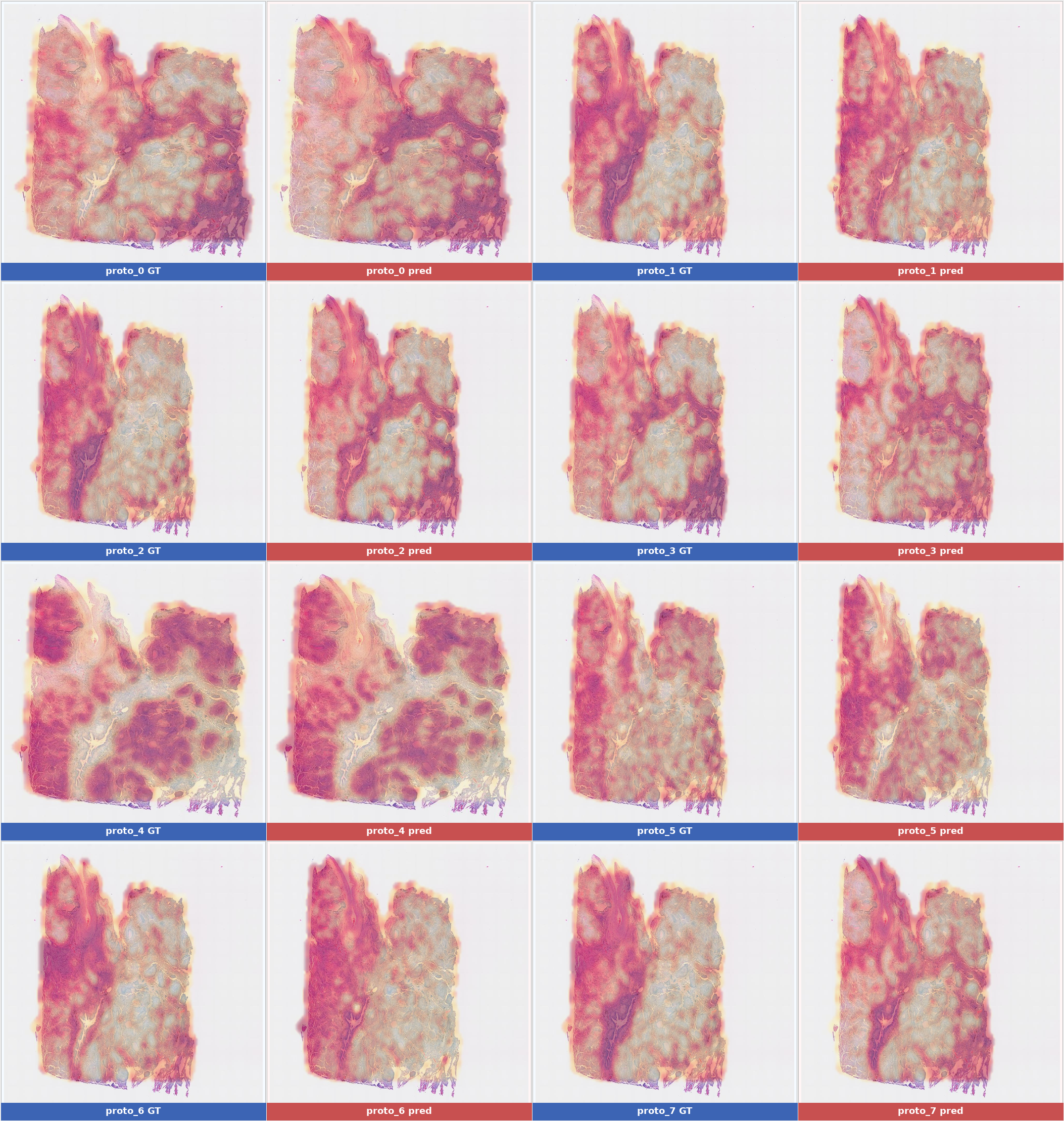}
    \caption{\textbf{Additional interpretability results on a held-out lung HEST slide.} Whole-slide prototype affinity visualizations generated from H\&E images without transcriptomics at inference.}
    \label{fig:interp_lung}
\end{figure*}

\section{Computational Cost}
\label{app:cost}

\begin{figure}[t]
    \centering
    \includegraphics[width=\linewidth]{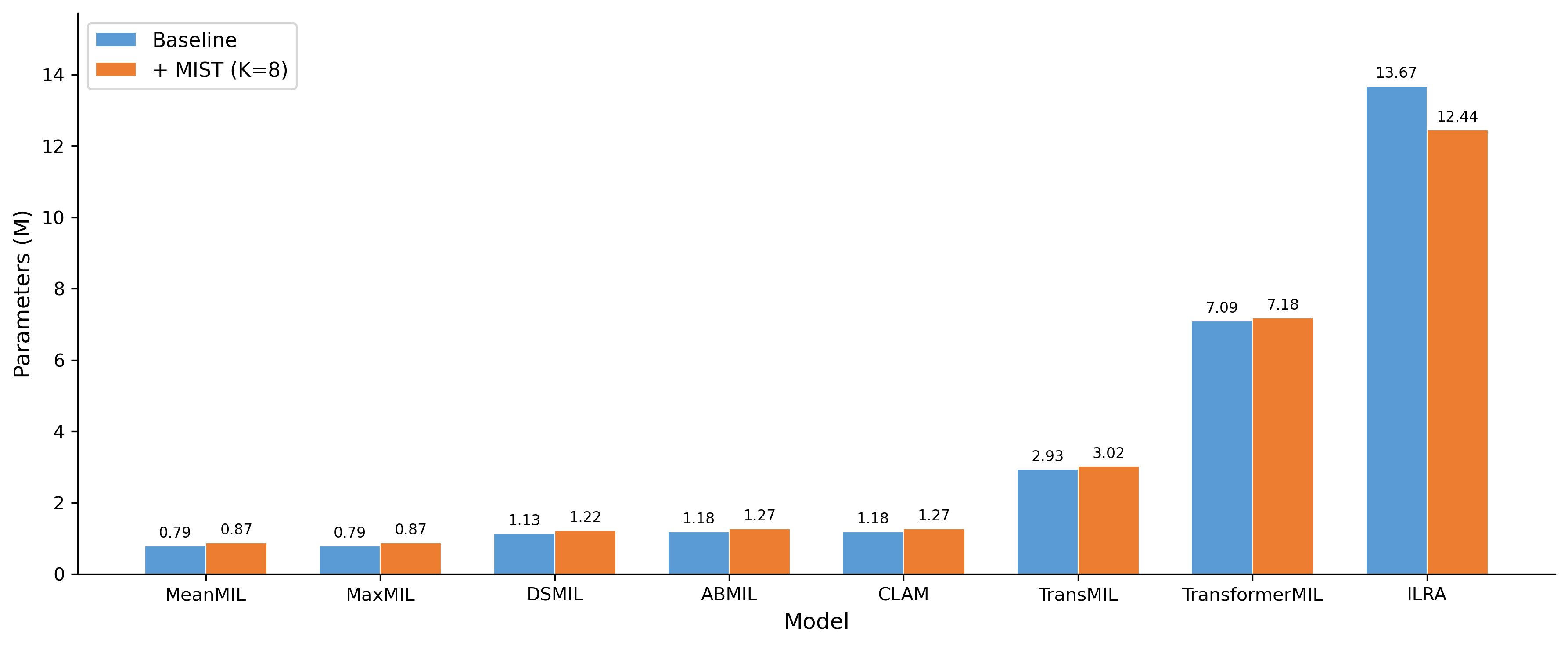}
    \caption{\textbf{Parameter comparison before and after inserting MIST.} Parameter counts (millions) across eight MIL aggregators with and without MIST ($K{=}8$, $r{=}16$). MIST introduces minimal overhead for most architectures and reduces the parameter count for ILRA through shared feature compression.}
    \label{fig:param_cost}
\end{figure}

MIST is designed as a lightweight plug-and-play module with minimal computational overhead. Figure~\ref{fig:param_cost} compares parameter counts before and after inserting MIST across eight MIL aggregators. With $K{=}8$ prototypes and rank $r{=}16$, the entire MIST module contains only $0.87$M parameters, most of which ($0.79$M) belong to the shared projection layer replacing the baseline input projection. As a result, the net parameter increase is less than $0.09$M for seven of the eight evaluated architectures. For example, ABMIL increases from $1.18$M to $1.27$M parameters ($+7.2\%$), while the overhead remains small for larger transformer-based models such as TransformerMIL ($7.09$M $\rightarrow$ $7.18$M, $+1.2\%$) and TransMIL ($2.93$M $\rightarrow$ $3.02$M, $+2.9\%$).

Interestingly, MIST reduces the parameter count of ILRA from $13.67$M to $12.44$M ($-9.0\%$). This reduction arises because the shared projection compresses the input feature dimension from $1{,}536$ to $512$ before the first attention block, reducing the size of subsequent projection layers. The efficiency of MIST further comes from the low-rank factorization of prototype-conditioned transforms, where each prototype contributes only a shared $\mathbf{V} \in \mathbb{R}^{D \times r}$ and a prototype-specific $\mathbf{U}_k \in \mathbb{R}^{r \times D}$ with small rank $r{=}16$, avoiding the cost of multiple full-rank transformations. Overall, these results show that MIST introduces spatial transcriptomic priors with minimal additional computational cost.

\section{Generalization Across Foundation Models}
\label{app:conch}

We repeat representative experiments with CONCH~\cite{lu2024visual} as the frozen encoder, re-encoding the HEST patches and rebuilding the prototype bank in its feature space. Results are reported in Table~\ref{tab:conch}.

\begin{table}[h]
\centering
\caption{\textbf{MIST with CONCH as the frozen encoder.} BRCA survival (C-index), BRCA biomarker prediction (mean over 4 tasks), and NSCLC subtyping (AUROC) using eight MIL aggregators.}
\label{tab:conch}
\resizebox{\textwidth}{!}{%
\begin{tabular}{llccccccccc}
\toprule
Task & Status & ABMIL & CLAM & TransMIL & Trans. & ILRA & Mean & Max & DSMIL & Avg. \\
\midrule
\multirow{3}{*}{BRCA survival}
 & Base  & 66.80 & 66.91 & 56.30 & 58.79 & 58.80 & 68.80 & 56.97 & 66.62 & 62.50 \\
 & +Ours & 68.57 & 68.84 & 60.97 & 64.34 & 68.08 & 68.27 & 61.26 & 67.57 & 65.99 \\
 & $\Delta$ & \textcolor{gaingreen}{+1.77} & \textcolor{gaingreen}{+1.94} & \textcolor{gaingreen}{+4.67} & \textcolor{gaingreen}{+5.55} & \textcolor{gaingreen}{+9.27} & \textcolor{lossred}{-0.53} & \textcolor{gaingreen}{+4.30} & \textcolor{gaingreen}{+0.96} & \textcolor{gaingreen}{+3.49} \\
\midrule
\multirow{3}{*}{\shortstack[l]{BRCA biomarker\\(4 tasks)}}
 & Base  & 74.93 & 74.88 & 72.68 & 73.61 & 70.60 & 71.98 & 66.07 & 72.41 & 72.15 \\
 & +Ours & 75.94 & 76.22 & 74.42 & 75.53 & 72.72 & 73.56 & 67.87 & 73.92 & 73.77 \\
 & $\Delta$ & \textcolor{gaingreen}{+1.01} & \textcolor{gaingreen}{+1.34} & \textcolor{gaingreen}{+1.74} & \textcolor{gaingreen}{+1.92} & \textcolor{gaingreen}{+2.12} & \textcolor{gaingreen}{+1.58} & \textcolor{gaingreen}{+1.80} & \textcolor{gaingreen}{+1.51} & \textcolor{gaingreen}{+1.63} \\
\midrule
\multirow{3}{*}{NSCLC}
 & Base  & 98.01 & 98.01 & 97.65 & 97.94 & 96.90 & 96.87 & 97.01 & 97.76 & 97.52 \\
 & +Ours & 98.00 & 98.01 & 97.89 & 98.03 & 97.50 & 97.02 & 97.13 & 98.08 & 97.71 \\
 & $\Delta$ & \textcolor{lossred}{-0.01} & 0.00 & \textcolor{gaingreen}{+0.24} & \textcolor{gaingreen}{+0.09} & \textcolor{gaingreen}{+0.59} & \textcolor{gaingreen}{+0.15} & \textcolor{gaingreen}{+0.12} & \textcolor{gaingreen}{+0.32} & \textcolor{gaingreen}{+0.19} \\
\bottomrule
\end{tabular}}
\end{table}

\section{MAMMOTH with an H\&E-only Control}
\label{app:mammoth}

Table~\ref{tab:mammoth_control} extends Table~\ref{tab:luad_orthogonal} with an H\&E-only control, which replaces ST-derived prototypes with morphology-only clusters. Results on PIK3CA prediction in BRCA are reported in Table~\ref{tab:biomarker_detailed}.

\begin{table}[h]
\centering
\caption{\textbf{MIST with MAMMOTH and an H\&E-only control.} LUAD survival (c-index).}
\label{tab:mammoth_control}
\small
\begin{tabular}{@{}llccc@{}}
\toprule
Task & Model & w/o MIST & + H\&E-only & + MIST \\
\midrule
LUAD   & ABMIL + MAMMOTH & 0.6324 & 0.6272 & 0.6647 \\
\bottomrule
\end{tabular}
\end{table}

\section{Limitations}

MIST relies on paired ST and H\&E data for prototype construction, which limits the diversity of molecular concepts to the tissues and cohorts currently available in existing ST datasets. Although the learned prototype banks generalize across multiple downstream datasets and external cohorts, they may not fully capture rare tissue states, treatment-specific phenotypes, or underrepresented patient populations.

\textbf{Choice of the projection layer.} Injecting ST supervision at the projection layer avoids retraining the foundation model and keeps MIST compatible with existing MIL aggregators. Jointly training a pathology--ST foundation model may be more powerful, but requires substantially more data and compute.

\textbf{Scope of the molecular contribution.} MIST reorganizes frozen H\&E features and cannot recover molecular variation that leaves no morphological trace; its contribution beyond morphology-only prototypes is modest but identifiable.

\textbf{Encoder specificity and interpretability.} The prototype bank must be rebuilt for each foundation model, and spatial interpretability is evaluated on one held-out HEST slide per organ.

\section{Broader Impacts}

MIST aims to improve whole-slide image analysis by introducing molecular priors into pathology foundation model pipelines without requiring transcriptomic measurements at inference. This may improve biomarker prediction, prognosis estimation, and tissue characterization while reducing dependence on expensive molecular assays. Because MIST operates as a lightweight plug-in module, it may also broaden access to molecularly-informed computational pathology systems in settings where transcriptomics is unavailable.

At the same time, MIST relies on molecular supervision derived from existing ST datasets, which remain limited in scale, tissue diversity, and population coverage. As a result, biases or underrepresentation in current ST datasets may propagate into downstream predictions and affect model generalization across institutions or patient populations.


\clearpage
\section*{NeurIPS Paper Checklist}

\begin{enumerate}

\item {\bf Claims}
    \item[] Question: Do the main claims made in the abstract and introduction accurately reflect the paper's contributions and scope?
    \item[] Answer: \answerYes{} 
    \item[] Justification: The abstract and introduction accurately describe the paper's scope and contributions.
    \item[] Guidelines:
    \begin{itemize}
        \item The answer \answerNA{} means that the abstract and introduction do not include the claims made in the paper.
        \item The abstract and/or introduction should clearly state the claims made, including the contributions made in the paper and important assumptions and limitations. A \answerNo{} or \answerNA{} answer to this question will not be perceived well by the reviewers. 
        \item The claims made should match theoretical and experimental results, and reflect how much the results can be expected to generalize to other settings. 
        \item It is fine to include aspirational goals as motivation as long as it is clear that these goals are not attained by the paper. 
    \end{itemize}

\item {\bf Limitations}
    \item[] Question: Does the paper discuss the limitations of the work performed by the authors?
    \item[] Answer: \answerYes{} 
    \item[] Justification: The paper discusses several limitations of MIST, including its reliance on paired spatial transcriptomics datasets for prototype construction and the limited diversity of current ST datasets. The paper also notes that rare tissue states and underrepresented patient populations may not be fully captured by the learned prototype banks.
    \item[] Guidelines:
    \begin{itemize}
        \item The answer \answerNA{} means that the paper has no limitation while the answer \answerNo{} means that the paper has limitations, but those are not discussed in the paper. 
        \item The authors are encouraged to create a separate ``Limitations'' section in their paper.
        \item The paper should point out any strong assumptions and how robust the results are to violations of these assumptions (e.g., independence assumptions, noiseless settings, model well-specification, asymptotic approximations only holding locally). The authors should reflect on how these assumptions might be violated in practice and what the implications would be.
        \item The authors should reflect on the scope of the claims made, e.g., if the approach was only tested on a few datasets or with a few runs. In general, empirical results often depend on implicit assumptions, which should be articulated.
        \item The authors should reflect on the factors that influence the performance of the approach. For example, a facial recognition algorithm may perform poorly when image resolution is low or images are taken in low lighting. Or a speech-to-text system might not be used reliably to provide closed captions for online lectures because it fails to handle technical jargon.
        \item The authors should discuss the computational efficiency of the proposed algorithms and how they scale with dataset size.
        \item If applicable, the authors should discuss possible limitations of their approach to address problems of privacy and fairness.
        \item While the authors might fear that complete honesty about limitations might be used by reviewers as grounds for rejection, a worse outcome might be that reviewers discover limitations that aren't acknowledged in the paper. The authors should use their best judgment and recognize that individual actions in favor of transparency play an important role in developing norms that preserve the integrity of the community. Reviewers will be specifically instructed to not penalize honesty concerning limitations.
    \end{itemize}

\item {\bf Theory assumptions and proofs}
    \item[] Question: For each theoretical result, does the paper provide the full set of assumptions and a complete (and correct) proof?
    \item[] Answer: \answerYes{} 
    \item[] Justification: The paper states the assumptions underlying the theoretical analysis in the main text and provides complete proofs in the appendix. In particular, the decomposition showing how prototype-dependent routing can increase separation between morphologically similar but molecularly distinct patches is formally derived and proved in Appendix
    \item[] Guidelines:
    \begin{itemize}
        \item The answer \answerNA{} means that the paper does not include theoretical results. 
        \item All the theorems, formulas, and proofs in the paper should be numbered and cross-referenced.
        \item All assumptions should be clearly stated or referenced in the statement of any theorems.
        \item The proofs can either appear in the main paper or the supplemental material, but if they appear in the supplemental material, the authors are encouraged to provide a short proof sketch to provide intuition. 
        \item Inversely, any informal proof provided in the core of the paper should be complemented by formal proofs provided in appendix or supplemental material.
        \item Theorems and Lemmas that the proof relies upon should be properly referenced. 
    \end{itemize}

    \item {\bf Experimental result reproducibility}
    \item[] Question: Does the paper fully disclose all the information needed to reproduce the main experimental results of the paper to the extent that it affects the main claims and/or conclusions of the paper (regardless of whether the code and data are provided or not)?
    \item[] Answer: \answerYes{} 
    \item[] Justification: The paper provides detailed implementation and experimental settings, including dataset descriptions, preprocessing procedures, model architectures, optimization settings, hyperparameters, evaluation metrics, and baseline configurations.
    \item[] Guidelines:
    \begin{itemize}
        \item The answer \answerNA{} means that the paper does not include experiments.
        \item If the paper includes experiments, a \answerNo{} answer to this question will not be perceived well by the reviewers: Making the paper reproducible is important, regardless of whether the code and data are provided or not.
        \item If the contribution is a dataset and\slash or model, the authors should describe the steps taken to make their results reproducible or verifiable. 
        \item Depending on the contribution, reproducibility can be accomplished in various ways. For example, if the contribution is a novel architecture, describing the architecture fully might suffice, or if the contribution is a specific model and empirical evaluation, it may be necessary to either make it possible for others to replicate the model with the same dataset, or provide access to the model. In general. releasing code and data is often one good way to accomplish this, but reproducibility can also be provided via detailed instructions for how to replicate the results, access to a hosted model (e.g., in the case of a large language model), releasing of a model checkpoint, or other means that are appropriate to the research performed.
        \item While NeurIPS does not require releasing code, the conference does require all submissions to provide some reasonable avenue for reproducibility, which may depend on the nature of the contribution. For example
        \begin{enumerate}
            \item If the contribution is primarily a new algorithm, the paper should make it clear how to reproduce that algorithm.
            \item If the contribution is primarily a new model architecture, the paper should describe the architecture clearly and fully.
            \item If the contribution is a new model (e.g., a large language model), then there should either be a way to access this model for reproducing the results or a way to reproduce the model (e.g., with an open-source dataset or instructions for how to construct the dataset).
            \item We recognize that reproducibility may be tricky in some cases, in which case authors are welcome to describe the particular way they provide for reproducibility. In the case of closed-source models, it may be that access to the model is limited in some way (e.g., to registered users), but it should be possible for other researchers to have some path to reproducing or verifying the results.
        \end{enumerate}
    \end{itemize}

\item {\bf Open access to data and code}
    \item[] Question: Does the paper provide open access to the data and code, with sufficient instructions to faithfully reproduce the main experimental results, as described in supplemental material?
    \item[] Answer: \answerYes{} 
    \item[] Justification: The code, prototype construction pipeline, and instructions to reproduce the main experimental results are publicly available at \url{https://github.com/YuchengXing99/MIST}. All datasets used are publicly available, and the paper provides detailed experimental settings, implementation details, and dataset descriptions.
    \item[] Guidelines:
    \begin{itemize}
        \item The answer \answerNA{} means that paper does not include experiments requiring code.
        \item Please see the NeurIPS code and data submission guidelines (\url{https://neurips.cc/public/guides/CodeSubmissionPolicy}) for more details.
        \item While we encourage the release of code and data, we understand that this might not be possible, so \answerNo{} is an acceptable answer. Papers cannot be rejected simply for not including code, unless this is central to the contribution (e.g., for a new open-source benchmark).
        \item The instructions should contain the exact command and environment needed to run to reproduce the results. See the NeurIPS code and data submission guidelines (\url{https://neurips.cc/public/guides/CodeSubmissionPolicy}) for more details.
        \item The authors should provide instructions on data access and preparation, including how to access the raw data, preprocessed data, intermediate data, and generated data, etc.
        \item The authors should provide scripts to reproduce all experimental results for the new proposed method and baselines. If only a subset of experiments are reproducible, they should state which ones are omitted from the script and why.
        \item At submission time, to preserve anonymity, the authors should release anonymized versions (if applicable).
        \item Providing as much information as possible in supplemental material (appended to the paper) is recommended, but including URLs to data and code is permitted.
    \end{itemize}

\item {\bf Experimental setting/details}
    \item[] Question: Does the paper specify all the training and test details (e.g., data splits, hyperparameters, how they were chosen, type of optimizer) necessary to understand the results?
    \item[] Answer: \answerYes{} 
    \item[] Justification: The paper specifies the datasets, preprocessing pipeline, feature extraction procedure, MIL aggregators, optimization settings, learning schedules, hyperparameters, evaluation metrics, and training protocols used in all experiments. Additional implementation details and ablation settings are provided in the appendix.
    \item[] Guidelines:
    \begin{itemize}
        \item The answer \answerNA{} means that the paper does not include experiments.
        \item The experimental setting should be presented in the core of the paper to a level of detail that is necessary to appreciate the results and make sense of them.
        \item The full details can be provided either with the code, in appendix, or as supplemental material.
    \end{itemize}

\item {\bf Experiment statistical significance}
    \item[] Question: Does the paper report error bars suitably and correctly defined or other appropriate information about the statistical significance of the experiments?
    \item[] Answer: \answerYes{} 
    \item[] Justification: The paper reports propagated standard errors in all main result tables. Survival, biomarker, and NSCLC tasks report standard errors across matched cross-validation folds; PANDA, BRACS, and EBRAINS report standard errors from 1000 bootstrap resamples of the test set. Results are consistently reported across multiple datasets, tasks, and MIL aggregators.
    \item[] Guidelines:
    \begin{itemize}
        \item The answer \answerNA{} means that the paper does not include experiments.
        \item The authors should answer \answerYes{} if the results are accompanied by error bars, confidence intervals, or statistical significance tests, at least for the experiments that support the main claims of the paper.
        \item The factors of variability that the error bars are capturing should be clearly stated (for example, train/test split, initialization, random drawing of some parameter, or overall run with given experimental conditions).
        \item The method for calculating the error bars should be explained (closed form formula, call to a library function, bootstrap, etc.)
        \item The assumptions made should be given (e.g., Normally distributed errors).
        \item It should be clear whether the error bar is the standard deviation or the standard error of the mean.
        \item It is OK to report 1-sigma error bars, but one should state it. The authors should preferably report a 2-sigma error bar than state that they have a 96\% CI, if the hypothesis of Normality of errors is not verified.
        \item For asymmetric distributions, the authors should be careful not to show in tables or figures symmetric error bars that would yield results that are out of range (e.g., negative error rates).
        \item If error bars are reported in tables or plots, the authors should explain in the text how they were calculated and reference the corresponding figures or tables in the text.
    \end{itemize}

\item {\bf Experiments compute resources}
    \item[] Question: For each experiment, does the paper provide sufficient information on the computer resources (type of compute workers, memory, time of execution) needed to reproduce the experiments?
    \item[] Answer: \answerYes{} 
    \item[] Justification: The paper provides the main implementation details required to reproduce the experiments, including preprocessing resolution, patch size, encoder architecture, optimization settings, batch size, training schedule, and model dimensions. The appendix additionally reports the parameter overhead introduced by MIST across different MIL aggregators.
    \item[] Guidelines:
    \begin{itemize}
        \item The answer \answerNA{} means that the paper does not include experiments.
        \item The paper should indicate the type of compute workers CPU or GPU, internal cluster, or cloud provider, including relevant memory and storage.
        \item The paper should provide the amount of compute required for each of the individual experimental runs as well as estimate the total compute. 
        \item The paper should disclose whether the full research project required more compute than the experiments reported in the paper (e.g., preliminary or failed experiments that didn't make it into the paper). 
    \end{itemize}
    
\item {\bf Code of ethics}
    \item[] Question: Does the research conducted in the paper conform, in every respect, with the NeurIPS Code of Ethics \url{https://neurips.cc/public/EthicsGuidelines}?
    \item[] Answer: \answerYes{} 
    \item[] Justification: The authors are not aware of any aspect of the work that violates the NeurIPS Code of Ethics. The research uses publicly available or appropriately licensed datasets and is intended to support computational pathology research rather than replace expert clinical decision-making.
    \item[] Guidelines:
    \begin{itemize}
        \item The answer \answerNA{} means that the authors have not reviewed the NeurIPS Code of Ethics.
        \item If the authors answer \answerNo, they should explain the special circumstances that require a deviation from the Code of Ethics.
        \item The authors should make sure to preserve anonymity (e.g., if there is a special consideration due to laws or regulations in their jurisdiction).
    \end{itemize}

\item {\bf Broader impacts}
    \item[] Question: Does the paper discuss both potential positive societal impacts and negative societal impacts of the work performed?
    \item[] Answer: \answerYes{} 
    \item[] Justification: The paper discusses potential positive impacts of improving molecularly-informed computational pathology without requiring transcriptomics at inference, which may broaden accessibility in resource-limited settings. It also discusses potential negative impacts arising from biases and underrepresentation in current spatial transcriptomics datasets.
    \item[] Guidelines:
    \begin{itemize}
        \item The answer \answerNA{} means that there is no societal impact of the work performed.
        \item If the authors answer \answerNA{} or \answerNo, they should explain why their work has no societal impact or why the paper does not address societal impact.
        \item Examples of negative societal impacts include potential malicious or unintended uses (e.g., disinformation, generating fake profiles, surveillance), fairness considerations (e.g., deployment of technologies that could make decisions that unfairly impact specific groups), privacy considerations, and security considerations.
        \item The conference expects that many papers will be foundational research and not tied to particular applications, let alone deployments. However, if there is a direct path to any negative applications, the authors should point it out. For example, it is legitimate to point out that an improvement in the quality of generative models could be used to generate Deepfakes for disinformation. On the other hand, it is not needed to point out that a generic algorithm for optimizing neural networks could enable people to train models that generate Deepfakes faster.
        \item The authors should consider possible harms that could arise when the technology is being used as intended and functioning correctly, harms that could arise when the technology is being used as intended but gives incorrect results, and harms following from (intentional or unintentional) misuse of the technology.
        \item If there are negative societal impacts, the authors could also discuss possible mitigation strategies (e.g., gated release of models, providing defenses in addition to attacks, mechanisms for monitoring misuse, mechanisms to monitor how a system learns from feedback over time, improving the efficiency and accessibility of ML).
    \end{itemize}
    
\item {\bf Safeguards}
    \item[] Question: Does the paper describe safeguards that have been put in place for responsible release of data or models that have a high risk for misuse (e.g., pre-trained language models, image generators, or scraped datasets)?
    \item[] Answer: \answerNA{} 
    \item[] Justification: The paper does not release high-risk generative models, scraped datasets, or systems with substantial misuse potential. The work focuses on computational pathology representation learning using existing research datasets.
    \item[] Guidelines:
    \begin{itemize}
        \item The answer \answerNA{} means that the paper poses no such risks.
        \item Released models that have a high risk for misuse or dual-use should be released with necessary safeguards to allow for controlled use of the model, for example by requiring that users adhere to usage guidelines or restrictions to access the model or implementing safety filters. 
        \item Datasets that have been scraped from the Internet could pose safety risks. The authors should describe how they avoided releasing unsafe images.
        \item We recognize that providing effective safeguards is challenging, and many papers do not require this, but we encourage authors to take this into account and make a best faith effort.
    \end{itemize}

\item {\bf Licenses for existing assets}
    \item[] Question: Are the creators or original owners of assets (e.g., code, data, models), used in the paper, properly credited and are the license and terms of use explicitly mentioned and properly respected?
    \item[] Answer: \answerYes{} 
    \item[] Justification: The paper properly cites the datasets, pathology foundation models, and baseline methods used throughout the experiments, including HEST, UNI2-h, and prior MIL architectures. The work uses publicly available research assets under their respective terms of use.
    \item[] Guidelines:
    \begin{itemize}
        \item The answer \answerNA{} means that the paper does not use existing assets.
        \item The authors should cite the original paper that produced the code package or dataset.
        \item The authors should state which version of the asset is used and, if possible, include a URL.
        \item The name of the license (e.g., CC-BY 4.0) should be included for each asset.
        \item For scraped data from a particular source (e.g., website), the copyright and terms of service of that source should be provided.
        \item If assets are released, the license, copyright information, and terms of use in the package should be provided. For popular datasets, \url{paperswithcode.com/datasets} has curated licenses for some datasets. Their licensing guide can help determine the license of a dataset.
        \item For existing datasets that are re-packaged, both the original license and the license of the derived asset (if it has changed) should be provided.
        \item If this information is not available online, the authors are encouraged to reach out to the asset's creators.
    \end{itemize}

\item {\bf New assets}
    \item[] Question: Are new assets introduced in the paper well documented and is the documentation provided alongside the assets?
    \item[] Answer: \answerNA{} 
    \item[] Justification: The paper does not introduce new public datasets, benchmarks, or software assets as part of the submission.
    \item[] Guidelines:
    \begin{itemize}
        \item The answer \answerNA{} means that the paper does not release new assets.
        \item Researchers should communicate the details of the dataset\slash code\slash model as part of their submissions via structured templates. This includes details about training, license, limitations, etc. 
        \item The paper should discuss whether and how consent was obtained from people whose asset is used.
        \item At submission time, remember to anonymize your assets (if applicable). You can either create an anonymized URL or include an anonymized zip file.
    \end{itemize}

\item {\bf Crowdsourcing and research with human subjects}
    \item[] Question: For crowdsourcing experiments and research with human subjects, does the paper include the full text of instructions given to participants and screenshots, if applicable, as well as details about compensation (if any)? 
    \item[] Answer: \answerNA{} 
    \item[] Justification: The paper does not involve crowdsourcing experiments or research conducted with human participants.
    \item[] Guidelines:
    \begin{itemize}
        \item The answer \answerNA{} means that the paper does not involve crowdsourcing nor research with human subjects.
        \item Including this information in the supplemental material is fine, but if the main contribution of the paper involves human subjects, then as much detail as possible should be included in the main paper. 
        \item According to the NeurIPS Code of Ethics, workers involved in data collection, curation, or other labor should be paid at least the minimum wage in the country of the data collector. 
    \end{itemize}

\item {\bf Institutional review board (IRB) approvals or equivalent for research with human subjects}
    \item[] Question: Does the paper describe potential risks incurred by study participants, whether such risks were disclosed to the subjects, and whether Institutional Review Board (IRB) approvals (or an equivalent approval/review based on the requirements of your country or institution) were obtained?
    \item[] Answer: \answerNA{} 
    \item[] Justification: The paper uses existing de-identified research datasets and does not involve direct interaction with human participants or new human subject data collection.
    \item[] Guidelines:
    \begin{itemize}
        \item The answer \answerNA{} means that the paper does not involve crowdsourcing nor research with human subjects.
        \item Depending on the country in which research is conducted, IRB approval (or equivalent) may be required for any human subjects research. If you obtained IRB approval, you should clearly state this in the paper. 
        \item We recognize that the procedures for this may vary significantly between institutions and locations, and we expect authors to adhere to the NeurIPS Code of Ethics and the guidelines for their institution. 
        \item For initial submissions, do not include any information that would break anonymity (if applicable), such as the institution conducting the review.
    \end{itemize}

\item {\bf Declaration of LLM usage}
    \item[] Question: Does the paper describe the usage of LLMs if it is an important, original, or non-standard component of the core methods in this research? Note that if the LLM is used only for writing, editing, or formatting purposes and does \emph{not} impact the core methodology, scientific rigor, or originality of the research, declaration is not required.
    \item[] Answer: \answerNA{} 
    \item[] Justification: Large language models are not used as part of the core methodology, experiments, or scientific claims of this work.
    \item[] Guidelines:
    \begin{itemize}
        \item The answer \answerNA{} means that the core method development in this research does not involve LLMs as any important, original, or non-standard components.
        \item Please refer to our LLM policy in the NeurIPS handbook for what should or should not be described.
    \end{itemize}

\end{enumerate}

\end{document}